\documentclass[5p,twocolumn]{elsarticle}

\usepackage{soul}
\usepackage{fancyhdr}

\usepackage{array} % to allow us to specify the width of the column with text centered

\usepackage{indentfirst}
\usepackage{amsmath}
\usepackage{amsfonts}
\usepackage{amssymb}
\usepackage{graphicx}
\usepackage{hyperref}
\usepackage{mathtools}
\usepackage{verbatim}
\usepackage{epstopdf}
\usepackage{subfigure}
\usepackage{latexsym}
\usepackage{dcolumn}
\usepackage{epsf}
\usepackage{float}
\usepackage[table]{xcolor}
\usepackage{multirow}
\usepackage{makecell}
\usepackage{color}
\usepackage{acronym}

\usepackage{bm}
\usepackage{appendix}

\usepackage[normalem]{ulem}
\usepackage{xcolor} % to use more color names

\usepackage{url}
\usepackage[utf8]{inputenc}
\biboptions{numbers,sort&compress}  % need this, transform citation [1,2,3,4,5,6,7] to [1-7]
\usepackage{times}
\usepackage{sectsty}
\usepackage{stackengine}
\usepackage{caption}
\allsectionsfont{\normalsize} % font size for sections, such as introduction.
\makeatletter
\def\ps@pprintTitle{%
 \let\@oddhead\@empty
 \let\@evenhead\@empty
 \def\@oddfoot{}%
 \let\@evenfoot\@oddfoot}
\makeatother

\newcommand{\shorttitle}{Reliability of energy landscape analysis of resting-state functional MRI data}

\begin{document}
\begin{frontmatter}
\title{Reliability of energy landscape analysis of resting-state functional MRI data}

\author{Pitambar Khanra$^1$,
Johan Nakuci$^2$,
Sarah Muldoon$^{1,4}$,
Takamitsu Watanabe$^3$,
Naoki Masuda$^{1,4,*}$}
%\cortext[cor]{corresponding author}
\address{$^1$ Department of Mathematics, State University of New York at Buffalo, Buffalo, USA.\\
$^2$ School of Psychology, Georgia Institute of Technology, Atlanta, USA\\
$^3$ International Research Centre for Neurointelligence, The University of Tokyo, Japan\\
$^4$ Institute for Artificial Intelligence and Data Science, State University of New York at Buffalo, Buffalo, USA\\
$*$corresponding author (naokimas@buffalo.edu)}

\begin{abstract}
Energy landscape analysis is a data-driven method to analyze multidimensional time series, including functional magnetic resonance imaging (fMRI) data. It has been shown to be a useful characterization of fMRI data in health and disease. It fits an Ising model to the data and captures the dynamics of the data as movement of a noisy ball constrained on the energy landscape derived from the estimated Ising model. In the present study, we examine test-retest reliability of the energy landscape analysis. To this end, we construct a permutation test that assesses whether or not indices characterizing the energy landscape are more consistent across different sets of scanning sessions from the same participant (i.e., within-participant reliability) than across different sets of sessions from different participants (i.e., between-participant reliability). We show that the energy landscape analysis has significantly higher within-participant than between-participant test-retest reliability with respect to four commonly used indices. We also show that a variational Bayesian method, which enables us to estimate energy landscapes tailored to each participant, displays comparable test-retest reliability to that using the conventional likelihood maximization method. The proposed methodology paves the way to perform individual-level energy landscape analysis for given data sets with a statistically controlled reliability.
\end{abstract}

\end{frontmatter}

\begin{acronym}
\section*{List of abbreviations}
  \acro{ATN}{attention network}
  \acro{CON}{cingulo-opercular network}
  \acro{DMN}{default mode network}
  \acro{EEG}{electroencephalogram}
  \acro{FPN}{fronto-parietal network}
  \acro{fMRI}{functional magnetic resonance imaging}
  \acro{HCP}{Human Connectome Project}
  \acro{ICA}{independent component analysis}
  \acro{ICC}{intraclass correlation coefficient}
  \acro{TSS}{total sum of squares}
  \acro{MEG}{magnetoencephalogram}
  \acro{MEM}{maximum entropy model}
  \acro{MSC}{Midnight Scan Club}
  \acro{ROI}{region of interest}
  \acro{SAN}{salience network}
  \acro{SMN}{somatosensory and motor network}
  \acro{VAN}{ventral attention network}
\end{acronym}

\section{Introduction}

Brain activity is dynamic and nonlinear in nature. Such nonlinear brain dynamics are considered to underlie many functions of the brain such as cognition, action, and learning~\cite{Deco_plos2008dynamic, Deco_NatRevNeuro2011emerging, Rabinovich_2012principles}, and mathematical modeling is widely accepted as a useful tool for simulating such brain dynamics on different scales \cite{Breakspear_CerebralCortex2006unifying, Deco_plos2008dynamic, Breakspear_FrontierNeuroscience2010generative, Deco_NatRevNeuro2011emerging, Rabinovich_2012principles, Sanz_Neuroinformatics2013virtual, Einevoll_NatureNeuroscience2013modelling}.
There are also many methods for analyzing empirical data of neural dynamics, including dynamic causal modeling~\cite{Friston_Neuroimage2003, Kiebel_CogNeuro2008dynamic}, functional network analysis~\cite{Sporns_2016networks, Bassett_NetNeuro2017network}, its dynamic variants~\cite{Hutchison_Neuroimage2013dynamic, Calhoun_Neuron2014chronnectome, Filippi_FrontierNeuro2019resting}, and hidden Markov models ~\cite{Baker_elife2014fast, Vidaurre_NatCom2018spontaneous, Ezaki_EJN2021modelling}.

Population-level inferences are a common practice for analyzing brain activity in empirical data. However, both the structure and dynamics of the brain vary even among healthy individuals, let alone among individuals belonging to a disease group due to the heterogeneity of the disease. Therefore, although population-level inferences increase the data size and often help us to reach statistically significant observations, they may yield inaccurate results and loss of information when the observed data are individual-specific. To avoid population-level inferences, it is necessary to establish the reliability of individual-level inferences of collective brain dynamics.

Finn et al.\,examined the role of individual variability in functional networks measured by functional magnetic resonance imaging (fMRI) and its ability to act as a fingerprint to identify individuals~\cite{Finn_NatNeuro2015functional} (see \cite{Anderson_Neuroradiology2011reproducibility, Miranda_PlosOne2014connectotyping} for earlier studies). In order for individual fingerprinting to be successful, the test-retest reliability of the functional network must be higher across sessions obtained from the same individual (i.e., within-participant reliability) than across sessions obtained from different individuals (i.e., between-participant reliability). Indeed, it was found that within-participant reliability was robust and that both resting-state and task fMRI from different sessions of the same individual could be used to perform fingerprinting~\cite{Noble_Neuroimage2019decade}.
Other studies also confirmed the ability of functional networks from fMRI data as fingerprints of individuals, including the development of different methods to quantify and improve fingerprinting~\cite{Gordon_Neuron2017precision, Amico_SciReport2018quest, Venkatesh_Neuroimage2020comparing, Chiem_Brain2022improving, Nakuci_SciReport2023within}. The ability of functional connectivity to act as individual fingerprints has also been confirmed with electroencephalogram (EEG)~\cite{Van_Brain2019test} and magnetoencephalogram (MEG) data~\cite{Garces_BrainConnec2016quantifying, Candelaria_Schizophrenia2020reduced}.

Functional networks or its dynamic variants are not the only tools for analyzing brain dynamics or fingerprinting individuals. One way to analyze fMRI or other multidimensional time series data from the brain is to infer dynamics of discrete states. Each state may correspond to a particular functional network~\cite{Allen_CerebralCortex2014tracking, Baker_elife2014fast, Calhoun_Neuron2014chronnectome, Ryali_Plos2016temporal, Taghia_Neuroimage2017bayesian, Nielsen_Neuroimage2018predictive} or a spatial activation pattern~\cite{Yan_PlosOne2009spontaneous, Rashid_FrontierNeuro2014dynamic, Ezaki_EJN2021modelling}, and the transition from one state to another may correspond to a regime shift in the brain. Energy landscape analysis is a method to characterize brain dynamics as a movement of a stochastic ball constrained on an energy landscape inferred from the data~\cite{Watanabe_Frontier2014energy, Watanabe_NatCom2014energy, Ezaki_PTRS2017energy}. Quantifications of the estimated energy landscapes such as the height of the barrier between two local minima of the energy allow intuitive interpretations; a local minimum of the energy is a particular spatial activity pattern and defines a discrete brain state. A high barrier between two local minima implies that it is difficult for the brain dynamics to transit between the two local minima. Indices from energy landscape analysis have been shown to be associated with behavior of healthy individuals in a test of bistable visual perception task~\cite{Watanabe_NatCom2014energy, Watanabe_Elife2021causal}, executive function~\cite{Kang_HBM2021bayesian}, fluid intelligence \cite{Ezaki_CommBio2020closer}, healthy aging~\cite{Ezaki_HBM2018}, autism~\cite{Watanabe_NatCom2017brain}, Alzheimer disease~\cite{Klepl_IEEE2021characterising}, schizophrenia~\cite{Allen_SPIE2021energy, Braun_NatComm2021brain}, attention deficit hyperactivity disorder~\cite{Udall_SmartBio2022comparing}, and epilepsy~\cite{Krzeminski_NetNeuro2020energy}.

These successful applications of energy landscape analysis are likely to owe to advantages of the method compared to other related methods such as functional network analysis and hidden Markov models. For example, with energy landscape analysis, one can borrow concepts and computational tools from statistical physics of spin systems to quantify the ease of state transition by the energy barrier~\cite{Ezaki_PTRS2017energy} and complexity of the dynamics by different phases (e.g., spin-glass phase) and susceptibility indices~\cite{Ezaki_CommBio2020closer}. In addition, each network state is by definition a binary activity pattern among a pre-specified set of regions of interest (ROIs) and therefore relatively easy to interpret. Despite its expanding applications, the validity of the energy landscape analysis has not been extensively studied except that one can measure the accuracy of fit of the model to the given data~\cite{Schneidman_PRL2003network, Schneidman_Nature2006weak, Shlens_JourNeuro2006structure, Yeh_Entropy2010maximum, Ezaki_PTRS2017energy}. A high accuracy of fit does not imply that the estimated energy landscape is a reliable fingerprint for individuals. In fact, if fMRI data are nonstationary, an energy landscape estimated for the same individual in two time windows may be substantially different from each other, whereas the accuracy of fit may be high in both time windows. Furthermore, the original energy landscape analysis method requires pooling of fMRI data from different individuals unless the number of regions of interest (ROIs) to be used is relatively small (e.g., 7) or the scanning session is extremely long. This is because the method is relatively data hungry~\cite{Ezaki_PTRS2017energy}. The concept of individual fingerprinting is unclear when pooling of data is necessary.

The present study is a precursor to being able to assess individual differences. We assess potential utility of energy landscape analysis in individual fingerprinting by investigating its test-retest reliability. Specifically, we ask how much features of the estimated energy landscapes are reproducible across different sessions from the same individual as opposed to across sessions belonging to different sets of individuals. We hypothesize that test-retest reliability is higher between sessions for the same individual than between sessions for different individuals. Code for computing energy landscapes with the conventional and Bayesian methods is available on Github~\cite{energy-landscape-analysis}.

\section{Methods}

\subsection{Midnight Scan Club data\label{sub:MSC-data}}

We primarily use the fMRI data in the Midnight Scan Club (MSC) data set~\cite{Gordon_Neuron2017precision}.
%
% The data are available at https://openneuro.org/datasets/ds000224/versions/00002. 
%
MSC data set contains five hours of resting-state fMRI data in total recorded from each of the 10 healthy human adults across 10 consecutive nights. A resting-state fMRI scanning section lasted for 30 minutes and yielded 818 volumes. Imaging was performed with a Siemens TRIO 3T MRI scanner using an echo planar imaging (EPI) sequence (TR $= 2.2$ s, TE $= 27$ ms, flip angle $= 90^{\circ}$, voxel size $= 4$ mm $\times$ 4 mm $\times$ 4 mm, 36 slices).

The original paper reported that the eighth participant (i.e., MSC08) fell asleep, showed frequent and prolonged eye closures, and had systematically large head motion, resulting in much less reliable data than those obtained from the other participants~\cite{Gordon_Neuron2017precision}. We also noticed that the accuracy of fitting the energy landscape, which we will explain in section~\ref{sub:accuracy-fit}, fluctuated considerably across the different sessions for the tenth participant (i.e., MSC10), suggesting unstable quality of the MSC10's data across sessions. Therefore, we excluded MSC08 and MSC10 from the analysis.

We used SPM12 (http://www.fil.ion.ucl.ac.uk/spm) to preprocess the resting-state fMRI data as follows: we first conducted realignment, unwraping, slice-timing correction, and normalization to a standard template (ICBM 152); then, we performed regression analyses to remove the effects of head motion, white matter signals, and cerebrospinal fluid signals; finally, we conducted band-pass temporal filtering (0.01--0.1 Hz).

We determined the ROIs of the whole-brain network using an atlas with 264 spherical ROIs whose coordinates were set in a previous study~\cite{Power_Neuron2011functional}.
We then removed 50 ROIs labelled `uncertain' or `subcortical', which left us with 214 ROIs.
The 214 ROIs were labeled either of the nine functionally different brain networks, i.e., 
auditory network, dorsal attention network (DAN), ventral attention network (VAN), cingulo-opercular network (CON), default mode network (DMN), fronto-parietal network (FPN), salience network (SAN), somatosensory and motor network (SMN), or visual network. We merged the DAN, VAN, and CON into an attention network (ATN) to reduce the number of observables from nine to seven, as we did in our previous study~\cite{Watanabe_NatCom2017brain}. This is due to the relatively short length of the data and the fact that energy landscape analysis requires sufficiently long data sets if working with 9 observables. In fact, the DAN, VAN, and CON are considered to be responsible for similar attention-related cognitive activity~\cite{Power_Neuron2011functional}, justifying the merge of the three systems into the ATN. We call the obtained $N=7$ dimensional time series of the fMRI signal the whole-brain network. We calculated the fMRI signal for each of the seven networks (i.e., ATN, auditory network, DMN, FPN, SAN, SMN, and visual network) by averaging the fMRI signal over the volumes in the sphere of radius 4 mm in the ROI and over all ROIs belonging to the network. 

In addition to the whole-brain network, we used a separate 30-ROI coordinate system~\cite{Fair_PCB2009functional} and determined the multi-ROI DMN and CON. We used a different parcellation system for the DMN and CON than the 264-ROI system used for the whole-brain network. It is because the former (i.e., 30-ROI) coordinate system provides much fewer ROIs for the DMN and CON than the 264-ROI system does, which is convenient for energy landscape analysis.  The original study identified 12 and 7 ROIs for the DMN and CON, respectively~\cite{Fair_PCB2009functional}. To reduce the dimension of the DMN, we averaged over each pair of the symmetrically located right- and left-hemisphere ROIs in the DMN into one observable. The symmetrized DMN, which we simply call the DMN, has eight ROIs because four ROIs (i.e., amPFC, vmPFC, pCC, and retro splen) in the original DMN are almost on the midline and therefore have not undergone the averaging between the right- and left-hemisphere ROIs~\cite{Ezaki_HBM2018}. For the CON, we used the original seven ROIs as the observables. Note that the whole-brain network contains the DMN and CON as single observables, whereas the DMN and CON we are introducing here are themselves systems containing $N=8$ and $N=7$ observables, respectively.

We denote the fMRI signal for the $i$th ROI at time $t$ by ${x_i^t}$ $(i=1,\dots,N; t=1,\dots,t_{\max})$, where $N$ is the number of ROIs, and $t_{\max}$ is the number of time points.  We then removed the global signals and transformed the signals into their $z$-values using $z_i^t=(x_i^t-m^t)/s^t$, where $m^t$ and $s^t$ represent the mean and standard deviation, respectively, of $x_i^t$ over the $N$ ROIs at time $t$; $m^t$ is the global signal~\cite{Murphy_Neuroimage2017global}. The global signal in resting-state functional MRI data is considered to be dominated by physiological noise mainly originating from the respiratory, scanner-related, and motion-related artifacts. Global signal removal improves various quality-control metrics, enhances the anatomical specificity of functional-connectivity patterns, and can increase the behavioral variance~\cite{Aquino_Neuroimage2020identifying, Li_SciReport2019topography}. The same or similar global signal removal was carried out in previous energy landscape studies~\cite{Ezaki_HBM2018, Ezaki_CommBio2020closer}.

\subsection{Human Connectome Project data}

For validation, we also analyzed fMRI data that were recorded from healthy human participants and shared as the S1200 data in the Human Connectome Project (HCP)~\cite{Van_Neuroimage2013wu}. In the data set, 1200 adults between $22$–$35$ years old underwent four sessions of 15-min EPI sequence with a 3T Siemens Connectome-Skyra (TR $= 0.72$ s, TE $= 33.1$ ms, $72$ slices, $2.0$ mm isotropic, field of view (FOV) $= 208 \times 180$ mm) and a T1-weighted sequence (TR $= 2.4$ s, TE $= 2.14$ ms, $0.7$ mm isotropic, FOV $= 224 \times 224$ mm). Here, we limited our analysis to those included in the 100 unrelated participant subset released by the HCP. We confirmed that all these 100 participants were among the subset of participants who completed diffusion weighted MRI as well as two resting-state fMRI scans.

The resting-state fMRI data of each participant are composed of two sessions, and each session is broken down into a Left-Right (LR) and Right-Left (RL) phases. We used data from participants with at least 1150 volumes in each of the four sessions after removing volumes with motion artifacts, which left us with 87 participants. For the 87 participants, we first removed the volumes with motion artifacts. Then, we used the last 1150 volumes in each session to remove possible effects of transient.

We used independent component analysis (ICA) to remove nuisance and motion signals~\cite{Glasser_Neuroimage2013minimal}. Furthermore, any volumes with frame displacement greater than 0.2 mm~\cite{Jenkinson_Neuroimage2002improved} were excised~\cite{Power_Neuroimage2012spurious} because the ICA-FIX pipeline has been found not to fully remove motion-related artifacts~\cite{Burgess_BrainConnectivity2016evaluation, Siegel_Cerebral2017data}. We standardized each voxel by subtracting the temporal mean. Lastly, global signal regression of the same form as that for the MSC data (see section~\ref{sub:MSC-data}) was used for removing remaining noise. 

In each volume, we averaged the fMRI signal over all the voxels within each ROI of the AAL atlas~\cite{Tzourio_Neuroimage2002automated}. Note that this atlas is composed of 116 ROIs. Then, we mapped each cortical ROI to either of the parcellation scheme from the Schaefer-100 atlas~\cite{Schaefer_CerebralCortex2018local}. System assignment was based on minimizing the Euclidian distance from the centroid of an ROI in the AAL to the corresponding centroid of an ROI in the Schaefer atlas. We removed 42 ROIs labeled `subcortical' or `cerebellar', which left us with 74 ROIs. These 74 ROIs were labelled either of the $N=7$ functionally different brain networks: control network, DMN, DAN, limbic network, salience/ventral attention network, somatomotor network, and visual network, altogether defining a whole-brain network.

\subsection{Fitting of the pairwise maximum entropy model}
\label{sub:fitting-pairwise-MEM}

To carry out energy landscape analysis, we fit the pairwise maximum entropy model (MEM), also known as the Ising model, to the preprocessed fMRI data in essentially the same manner as in previous studies~\cite{Watanabe_NatCom2013pairwise, Ezaki_PTRS2017energy}.

For each session, we first binarized $z_i^t$ for each $i$th ROI (with $i\in \{1, \ldots, N\}$) and time $t$ (with $t\in \{1, \ldots, t_{\max} \}$) using a threshold that we set to the time average of $z_i^t$. A computational study showed that binarization did not affect important information contained in originally continuous brain signals~\cite{Deco_Front2012anatomy}. We denote the binarized signal at the $i$th ROI and time $t$ by $\sigma_i^t$, which is either $+1$ or $-1$ corresponding to whether $z_i^t$ is larger or smaller than the threshold, respectively. The activity pattern of the entire network at time $t$ is described by the $N$-dimensional vector
\begin{equation}
V^t = [ \sigma_1^t, \ldots, \sigma_N^t ] \in \{-1,1\}^N.
\end{equation}
It should be noted that there are $2^N$ activity patterns in total, enumerated as $V_1$, $\ldots$, $V_{2^N}$. The empirical mean activity at the $i$th ROI is denoted by
\begin{equation}
\langle \sigma_i \rangle \equiv \frac{1}{t_{\max}} \sum_{t=1}^{t_{\max}} \sigma_i^t.
\end{equation}
The empirical mean pairwise joint activation for the $i$th and $j$th ROIs is defined by
\begin{equation}
\langle \sigma_i \sigma_j \rangle \equiv \frac{1}{t_{\max}} \sum_{t=1}^{t_{\max}} \sigma_i^t \sigma_j^t.
\end{equation}

The pairwise MEM maximizes the entropy of the distribution of activity patterns under the condition that $\langle \sigma_i \rangle$ and $\langle \sigma_i \sigma_j \rangle$ (with $1 \leq i \leq j \leq N)$ are the same between the estimated model and the empirical data. The resulting probability distribution of activity pattern $V = [\sigma_1, \ldots, \sigma_N]$, denoted by $P(V)$, obeys the Boltzmann distribution~\cite{Jaynes_PhyRev1957information} given by
\begin{equation}
P(V)=\frac{e^{-E(V)}}{\sum_{k=1}^{2^N} e^{-E(V_k)}},
\label{eq:P(V)}
\end{equation}
where $E(V)$ represents the energy of activity pattern $V$ given by
\begin{equation}
E(V)=-\sum_{i=1}^N h_i \sigma_i - \frac{1}{2} \sum_{i=1}^N \sum_{j=1}^N J_{ij} \sigma_i \sigma_j.
\label{eq:E(V)}
\end{equation}
In Eq.~\eqref{eq:E(V)}, the fitting parameter $h_i$ represents the tendency for the $i$th ROI to be active (i.e., $\sigma_i = +1$), and $J_{ij}$ quantifies the pairwise interaction between the $i$th and $j$th ROIs.

We denote the mean activity and mean pairwise activity from the estimated model by $\langle \sigma_i \rangle_{\rm m}$ and $\langle \sigma_i \sigma_j \rangle_{\rm m}$, respectively. By definition, we obtain
\begin{equation}
\langle \sigma_i \rangle_{\rm m}= \sum_{k=1}^{2^N} \sigma_i(V_k) P(V_k)
\end{equation}
and
\begin{equation}
\langle \sigma_i \sigma_j \rangle_{\rm m} = \sum_{k=1}^{2^N} \sigma_i(V_k) \sigma_j(V_k) P(V_k).
\end{equation}

We calculated $h_i$ and $J_{ij}$ by iteratively adjusting $\langle \sigma_i \rangle_{\rm m}$ and $\langle \sigma_i \sigma_j \rangle_{\rm m}$ towards the empirically values, i.e., $\langle \sigma_i \rangle$ and $\langle \sigma_i \sigma_j \rangle$, respectively, using a gradient ascent algorithm. The iteration scheme is given by
\begin{equation}
h_i^{\text{new}}=h_i^{\text{old}} + \epsilon \log \frac{\langle \sigma_i \rangle} {\langle \sigma_i \rangle_{\rm m}}
\end{equation}
and
\begin{equation}
J_{ij}^{\text{new}}=J_{ij}^{\text{old}} + \epsilon \log \frac{\langle \sigma_i \sigma_j \rangle} {\langle \sigma_i \sigma_j \rangle_{\rm m}},
\end{equation}
where superscripts new and old represent the values after and before a single updating step, respectively, and $\epsilon$ is the learning rate. We set $\epsilon=0.2$.

\subsection{Accuracy of fit}\label{sub:accuracy-fit}

We evaluated the accuracy of fit of the pairwise MEM to the given fMRI data~\cite{Schneidman_Nature2006weak, Ezaki_PTRS2017energy, Ezaki_HBM2018}. The accuracy index is given by
\begin{equation}
r_D=\frac{D_1-D_2}{D_1},
\end{equation}
where
\begin{equation}
D_\ell=\sum_{k=1}^{2^N}P_N(V_k) \log_2 \frac{P_N(V_k)}{P_{\ell}(V_k)}
\label{eq:D_ell-def}
\end{equation}
is the Kullback-Leibler divergence between the probability distribution of the activity pattern in the $\ell$th-order $(\ell=1, 2)$ MEM, $P_{\ell}(V)$, and the empirical probability distribution of the activity pattern, denoted by $P_N(V)$. Note that $P_2(V)$ is equivalent to $P(V)$ given by Eqs.~\eqref{eq:P(V)} and \eqref{eq:E(V)}. The first-order, or independent, MEM (i.e., $\ell=1$) is Eq.~\eqref{eq:P(V)} estimated without interaction terms, that is, $J_{ij}=0$ $\forall i,j$ in Eq.~\eqref{eq:E(V)}. We obtain $r_D=1$ when the pairwise MEM perfectly fits the empirical distribution of the activity pattern, and $r_D=0$ when the pairwise MEM does not fit the data any better than the independent MEM.

To assess the dependency of $r_D$ on the number of sessions to be concatenated for the estimation of the pairwise MEM, $m$, the network (i.e., whole-brain, DMN, or CON), and the type of concatenation (i.e., within-participant or between-participant), we examined the multivariate linear regression model given by
\begin{equation}
r_D = \beta_0 + \beta_1 m + \beta_2 I_{\text{whole}} + \beta_3 I_{\text{CON}} + \beta_4 I_{\text{within}}.
\label{eq:r_D-regression}
\end{equation}
In Eq.~\eqref{eq:r_D-regression}, $\beta_0$ is the intercept, dummy variable $I_{\text{whole}}$ is equal to $1$ for the whole-brain network and 
$0$ for the other two networks, $I_{\text{CON}}$ is equal to $1$ for the CON, and $0$ for the other two networks, and $I_{\text{within}}$ is equal to $1$ for the within-participant comparison and $0$ for the between-participant comparison.

\subsection{Bayesian approximation method}

The pairwise MEM and the subsequent energy landscape analysis have mostly been restricted to analysis of group-level data. This is because the methods in its original form are data-hungry, requiring concatenation of fMRI signals from different individuals. If there are sufficiently many or long sessions of fMRI data from a single participant, as in the present study, one can only concatenate the data from the same participant and thus avoid group-level energy landscape analysis. However, fMRI data from a single participant are more often than not too short to allow individual-level energy landscape analysis. In addition, the length of fMRI data, $t_{\max}$, that is necessary for reliably estimating the pairwise MEM with $N$ nodes is roughly proportional to the number of states, $2^N$~\cite{Ezaki_PTRS2017energy}. To overcome this problem and obtain the energy landscape for each individual, we employed a recently developed variational Bayes approximation method for estimating the pairwise MEM~\cite{Kang_HBM2021bayesian,Jeong_Neuroimage2021empirical}, which runs as follows.

We denote by $\mathcal{S}_n$ the $N$-dimensional time series obtained from an $n$th session of fMRI. Different fMRI sessions typically originate from different participants in the same group (e.g., control group). We denote the number of sessions available by $D$.
Let $\mathcal{S}$ be the concatenated data, i.e.,
\begin{equation}
\mathcal{S} \equiv \cup_{n=1}^D \mathcal{S}_n.
\end{equation}
The variational Bayes approximation method estimates a pairwise MEM for each $\mathcal{S}_n$ (with $n\in \{1, \ldots, D\}$).

This method introduces a prior distribution for the set of session-specific model parameters, $\bm{\theta}_n =(h_1,h_2,\dots,h_N,J_{12},J_{13},\dots,J_{N-1,N})\in\mathbb R^M$, where $n \in \{ 1, \ldots, D \}$ and $M = N(N+1)/2$. We give the prior distribution for
\begin{equation}
\Theta=[\bm{\theta}_1, \ldots, \bm{\theta}_D]
\end{equation}
by
\begin{equation}
p(\Theta | \bm{\eta}, \bm{\alpha})=\prod_{n=1}^D \prod_{M'=1}^M p(\theta_{nM'} | \mathcal{N} (\eta_{M'}, 1/\alpha_{M'})),
\label{eq:prior-all-data}
\end{equation}
where $p(x | \mathcal{N} (\mu, \sigma^2))$ represents the probability density of $x$ obeying the one-dimensional normal distribution with mean and variance equal to $\mu$ and $\sigma^2$, respectively. Here, $\bm \eta=(\eta_1,\dots,\eta_M)^\top \in \mathbb R^M$ is the prior mean vector, $\bm \alpha=(\alpha_1,\dots,\alpha_M)^\top \in \mathbb R_+^M$ is the prior precision vector, and ${}^{\top}$ represents the transposition. In Eq.~\eqref{eq:prior-all-data}, we have assumed that the signals from all the $D$ sessions are mutually independent.

Now, we derive the posterior distribution of $\Theta$. It is intractable to derive the posterior because the normal distribution is not the conjugate prior for the Boltzmann distribution. Therefore, we use a variational approximation to the posterior~\cite{Bishop_book2006pattern} using the normal distribution as follows:
\begin{equation}
q(\Theta |\mathcal{S}, \bm{\eta}, \bm{\alpha}) = \prod_{n=1}^D \prod_{M'=1}^M p(\theta_{nM'} | \mathcal{N} (\mathbb \mu_{nM'}, 1/\beta_{nM'})).
\end{equation}

We write $\bm{\mu}_n=(\mu_{n1},\dots,\mu_{nM})^\top \in \mathbb R^M$ and $\bm{\beta}_n=(\beta_{n1},\dots,\beta_{nM})^\top \in \mathbb R_+^M$, which are the posterior mean vector and the posterior precision vector for session $n \in \{1, \ldots, D \}$, respectively.
One obtains the variational approximate solution for distribution $q$ by optimizing the evidence lower bound (ELBO), also called the free energy~\cite{Kang_HBM2021bayesian,Jeong_Neuroimage2021empirical}. By maximizing the free energy with respect to $q$, we have the posterior mean and precision vectors in terms of the prior mean and precision vectors as follows:
\begin{eqnarray}
\bm{\mu}_n &=& \bm{\eta} + t_{\max} \mathbb A_{\bm{\eta}, \bm{\alpha}}^{-1}(\langle \bar \sigma_n \rangle-\langle \bar \sigma \rangle_{\bm{\eta}}),\label{mu_n}\\
\bm{\beta}_n &=& \bm{\alpha} + t_{\max} \bm{c}_{\bm{\eta}},\label{beta_n}
\end{eqnarray}
where 
\begin{eqnarray}
\mathbb A_{\bm{\eta}, \bm{\alpha}} &=& \text{diag}(\bm{\alpha})+ t_{\max} {\rm C}_{\bm{\eta}},
\end{eqnarray}
and $\rm{diag}(\cdot)$ represents the diagonal matrix whose entries are given by the arguments.
In Eq.~\eqref{mu_n}, $\langle \bar \sigma_n \rangle \equiv (\langle \sigma_1 \rangle, \ldots, \langle \sigma_N\rangle, \langle \sigma_1 \sigma_2 \rangle, \langle \sigma_1 \sigma_3 \rangle, \ldots, \langle \sigma_{N-1} \sigma_N \rangle)^\top$ is the vector composed of the empirical mean activity and empirical pairwise joint activation; $\langle \bar \sigma \rangle_{\bm{\eta}}$ is the model mean of
$\bar \sigma_n \equiv  (\sigma_1,\sigma_2, \dots, \sigma_N, \sigma_1 \sigma_2, \sigma_1 \sigma_3, \dots, \sigma_{N-1} \sigma_N)^\top$
when the model parameters $(h_1,h_2,\dots,h_N,J_{12},J_{13},\dots,J_{N-1,N})$ are given by $\bm{\eta}$;
$\text{C}_{\bm{\eta}} \equiv \text{Cov}_{\bm{\eta}}(\bar \sigma_n)$ is the covariance matrix of $\bar{\sigma}_n$ when the model is given by $\bm{\eta}$.  
In Eq.~\eqref{beta_n}, $\bm{c}_{\bm{\eta}}$ is the vector composed of the diagonal element of $\text{C}_{\bm{\eta}}$.
In other words, the $i$th element of $\bm{c}_{\bm{\eta}}$ is the variance of the $i$th element of $\bar{\sigma}_n$ under parameters $\bm{\eta}$.

Now, we fix $q$ and maximize the free energy with respect to $\bm{\eta}$ and $\bm{\alpha}$ to obtain the equations for updating $\bm{\eta}$ and $\bm{\alpha}$ as follows:
\begin{eqnarray}
\eta_{M'} &=& \frac{1}{D}\sum_{n=1}^D \mathbb \mu_{nM'} \label{eta}, \\
\alpha_{M'} &=& \left[ \frac{1}{D} \sum_{n=1}^D \left\{(\mu_{nM'}-\eta_{M'})^2 + \frac{1}{\beta_{nM'}} \right\} \right]^{-1}, \label{alpha}
\end{eqnarray}
where $M' \in \{1, \ldots, M\}$.

Thus, we have updated the posterior distribution $\theta_{nM'} \sim \mathcal{N}(\mu_{nM'}, 1/\beta_{nM'})$, $n\in \{1, \ldots, D\}$, $M' \in \{1, \ldots, M\}$ using the prior distribution $\theta_{nM'} \sim \mathcal{N}(\eta_{M'}, 1/\alpha_{M'})$, and then updated the prior distribution using the new posterior distribution. We summarize the steps of the variational Bayes approximation method as follows:
\begin{enumerate}
\item Initialize the hyperparameters by independently drawing each $\eta_{M'}$ (with $M' \in \{1, \ldots, M\}$) from the normal distribution with mean $0$ and standard deviation $0.1$. We also set the first $N$ entries of the prior precision vector $\alpha_{M'}$, corresponding to $h_i$, $i\in \{1, \ldots, N\}$, to $6$, and
set the remaining $M-N$ entries of $\alpha_{M'}$ corresponding to $J_{ij}$, $1\le i < j \le N$, to $30$. 

\item Calculate the posterior mean vector and posterior precision vector for each $n \in \{ 1, \ldots, D\}$ using Eqs.~\eqref{mu_n} and \eqref{beta_n}.

\item Update the prior mean vector, $\bm{\eta} = (\eta_1, \ldots, \eta_M)^{\top}$, and the prior precision vector, $\bm{\alpha} = (\alpha_1, \ldots, \alpha_M)^{\top}$, using Eqs.~\eqref{eta} and \eqref{alpha}.

\item If $\left\lvert \frac{\rm{ELBO}(\rm{iter})}{\rm{ELBO}(\rm{iter}-1)}-1\right\rvert < 10^{-8}$, we stop the iteration. Otherwise, we return to step 2. Here, $\rm{ELBO}(\rm{iter})$ represents the ELBO value after `iter' iterations of steps 2 and 3.
\end{enumerate}

\subsection{Energy landscape and disconnectivity graph}

\begin{figure}[t]
\begin{center}
\includegraphics[height=!,width=8.3cm]{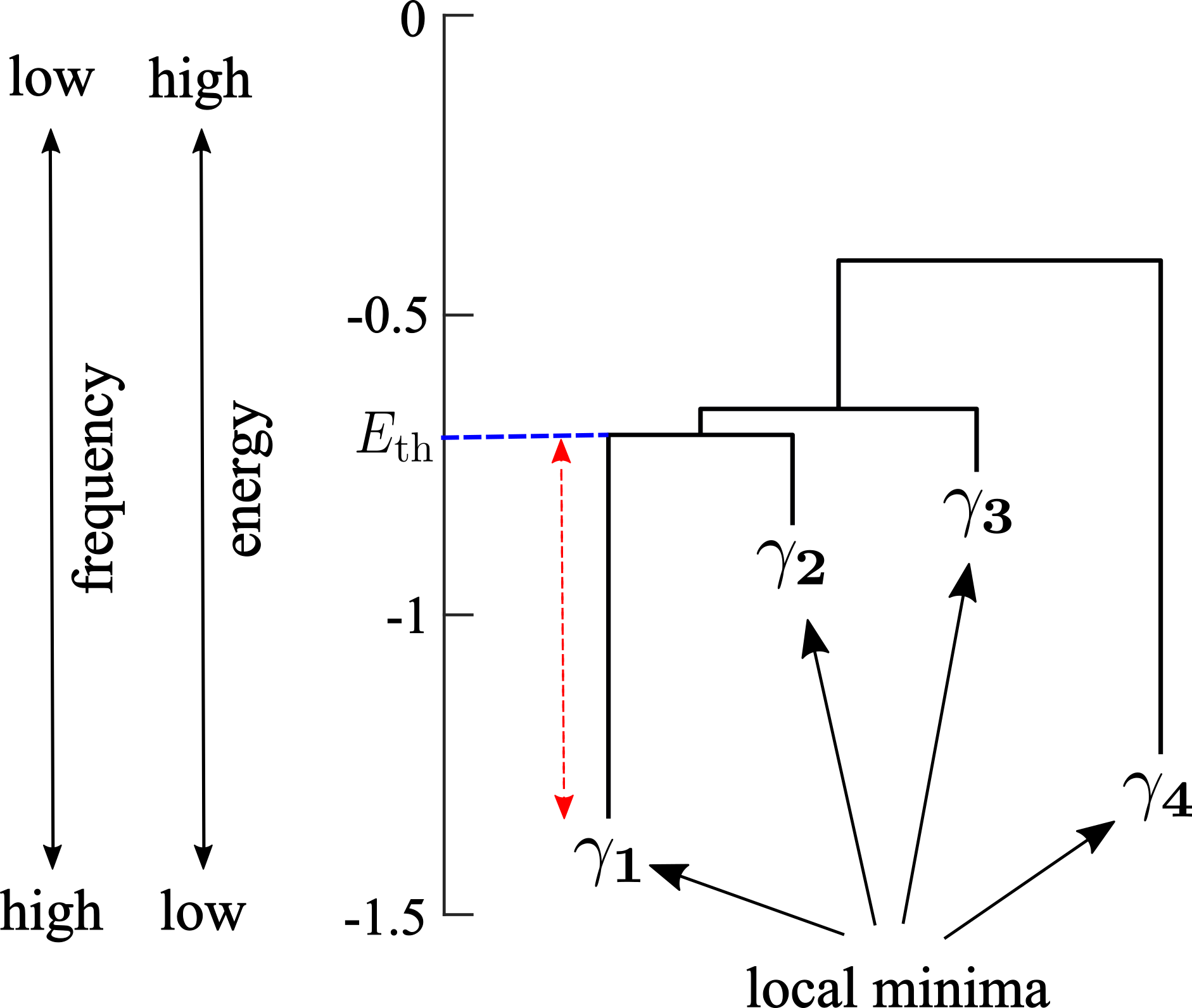}
\caption{Schematic of a disconnectivity graph showing the relationships between the activity patterns that are energy local minima. The arrow indicates the height of the energy barrier between local minima $\gamma_1$ and $\gamma_2$ from the viewpoint of $\gamma_1$.}
\label{fig:disconnectivity-graph-scehmatic}
\end{center}
\end{figure}

Once we have estimated the pairwise MEM, we calculated the energy landscape~\cite{Watanabe_NatCom2014energy, Watanabe_Frontier2014energy, Ezaki_PTRS2017energy, Ezaki_HBM2018}. The energy landscape is defined as a network with $2^N$ nodes in which each node is an activity pattern. We first constructed a dendrogram called the disconnectivity graph. We show a hypothetical disconnectivity graph in Fig.~\ref{fig:disconnectivity-graph-scehmatic}. 
See Appendix A~\ref{Sample_disconnectivity} for visualization of a real disconnectivity graph and its randomized counterpart.
In the disconnectivity graph, a leaf corresponds to an activity pattern $V_k$ that is a local minimum of the energy. 
There are four local minima in the disconnectivity graph shown in Fig.~\ref{fig:disconnectivity-graph-scehmatic}.
The vertical position of the leaf represents the energy value of the local minimum. A low energy value corresponds to a high frequency of appearance through Eq.~\eqref{eq:E(V)}. For example, in Fig.~\ref{fig:disconnectivity-graph-scehmatic}, activity pattern $\gamma_1$ is the one that appears with the highest frequency among all the $2^N$ activity patterns.

By definition, activity pattern $V_k$ is a local minimum of energy if and only if $V_k$ appears more frequently (thus has a lower energy) than any other activity pattern adjacent to $V_k$. Two activity patterns are defined to be adjacent in the network of activity patterns if and only if they have the opposite activity $\sigma_i \in \{ -1, 1 \}$ for just one $i$. Note that the network of activity patterns is the hypercube composed of $2^N$ nodes in which each node representing an activity pattern is adjacent to $N$ other nodes. To obtain the disconnectivity graph, we first enumerate the local minima. Then, for each pair of local minima $\gamma$ and $\gamma'$, we determine the smallest energy value $E_{\rm{th}}$ that a path connecting $\gamma$ and $\gamma'$ needs to go through as follows. There may be various paths connecting $\gamma$ and $\gamma'$. Then, we sequentially remove nodes in the descending order of the energy until there is no path connecting $\gamma$ and $\gamma'$. The energy of the node that we have removed the last is the $E_{\rm{th}}$ value for $\gamma$ and $\gamma'$. The horizontal dashed line in Fig.~\ref{fig:disconnectivity-graph-scehmatic} indicates the $E_{\rm{th}}$ value ($= -0.69$) for the pair of local minima $\gamma_1$ and $\gamma_2$. The difference between $E_{\rm{th}}$ and the energy at the local minimum represents the energy barrier that the dynamics of the brain have to overcome to reach from one local minima to the other. In Fig.~\ref{fig:disconnectivity-graph-scehmatic}, the energy barrier between $\gamma_1$ and $\gamma_2$ from the viewpoint of $\gamma_1$ is $0.64$, which is indicated by the double-headed arrow. The disconnectivity graph shows $E_{\rm{th}}$ and the energy barrier values for all pairs of the local minima.

\subsection{Measures of discrepancy}
\label{sub:discrepancy}

To assess within-participant test-retest reliability of energy landscape analysis, we compared two energy landscapes that we separately estimated for two sets of fMRI data, which were in different sessions for the same participant or obtained from different participants. We decided to make within-participant versus between-participant comparisons because a successful individual fingerprinting requires that the within-participant test-retest reliability is high enough, whose examination requires a baseline. Higher within-participant test-retest reliability than between-participant one implies that the energy landscape analysis provides reliable fingerprints for individuals. To analyze test-retest reliability, we measured the following four indices of the discrepancy between the two energy landscapes.

\subsubsection{Discrepancy in terms of the interaction strength}
\label{Jij}

The energy landscape is primarily a function of $\{ J_{ij} \}_{i, j \in \{1, \ldots, N\}}$ because $\{ h_1, \ldots, h_N \}$ tend to take values close to $0$ if we set our threshold to binarize $z_i^t$ such that the fraction of $\sigma_i = -1$ and that of $\sigma_i = 1$ are not heavily imbalanced~\cite{Ezaki_CommBio2020closer}. Therefore, we measured the discrepancy between two energy landscapes in terms of the estimated $\{ J_{ij} \}$. We define the discrepancy using the Frobenius distance as follows:
\begin{equation}
d_J = \frac{2}{N(N-1)}\sum_{i=1}^N \sum_{j=i+1}^N \left| J^{(1)}_{ij}-J^{(2)}_{ij} \right|,
\end{equation}
where $J^{(1)} = (J^{(1)}_{ij})$ and $J^{(2)} = (J^{(2)}_{ij})$ denote the pairwise interaction matrices according to the pairwise MEM estimated for the first and second data sets, respectively. 

\subsubsection{Discrepancy in terms of the activity patterns at the local minima of the energy}
\label{hamming_distance}

A local minimum of the energy landscape is locally the most frequent activity pattern. We compared the location of the local minima in the two energy landscapes by calculating the Hamming distance between the activity patterns at the local minima from the first energy landscape and those from the second energy landscape as follows.

First, we assumed that minor local minima characterized by low energy barriers with other local minima did not play important roles because the brain state would stay near such shallow local minima only briefly. Therefore, we started by removing minor local minima of the energy as follows. We generated $N$ random binary time series of length $4t_{\max}$ by independently drawing the $N\times 4t_{\max}$ binary numbers, i.e., $-1$ or $+1$, with the same probability (i.e., $0.5$). The multiplication factor was set at 4 because we mainly analyzed energy landscapes of the empirical fMRI data with $t_{\max}$ volumes that were concatenated over four sessions. Then, we inferred the pairwise MEM for the generated random binary time series and calculated the maximum length of the branch in the disconnectivity graph. A branch corresponds to a local minimum of the energy landscape. We define the branch length for local minimum $\gamma$ by the smallest value of the energy barrier between $\gamma$ and another local minimum $\gamma'$ among all local minima $\gamma' (\neq \gamma)$. In the disconnectivity graph shown in Fig.~\ref{fig:disconnectivity-graph-scehmatic}, the branch length for $\gamma_1$ is the length of the arrow. We claim that the energy landscape estimated for the random binary time series, including the number and depth of its local minima, does not have functional meanings. Therefore, in an energy landscape estimated for the empirical data, the local minima whose branch length is comparable with the maximum branch length for the random binary time series are not important. 

To implement this idea, we generated random binary time series, inferred the energy landscape, computed its maximum branch length, and repeated all these steps $100$ times. We denote the average and standard deviation of the maximum branch length on the basis of the 100 random binary time series by $\mu'$ and $\sigma'$, respectively. We then identified the local minimum with the shortest branch length in the original disconnectivity graph. We removed that local minimum as being insignificant if its branch was shorter than $\mu'+2\sigma'$. If we removed this local minimum, we recomputed the branch length of each local minimum whose branch had merged with the removed branch. Then, if the shortest branch was shorter than $\mu' + 2\sigma'$, we removed the branch and repeated these steps until all the local minima have branches whose length is at least $\mu' + 2\sigma'$. We refer to the local minima that survive this test as major local minima.

\begin{figure}[h]
\begin{center}
\includegraphics[width=8cm]{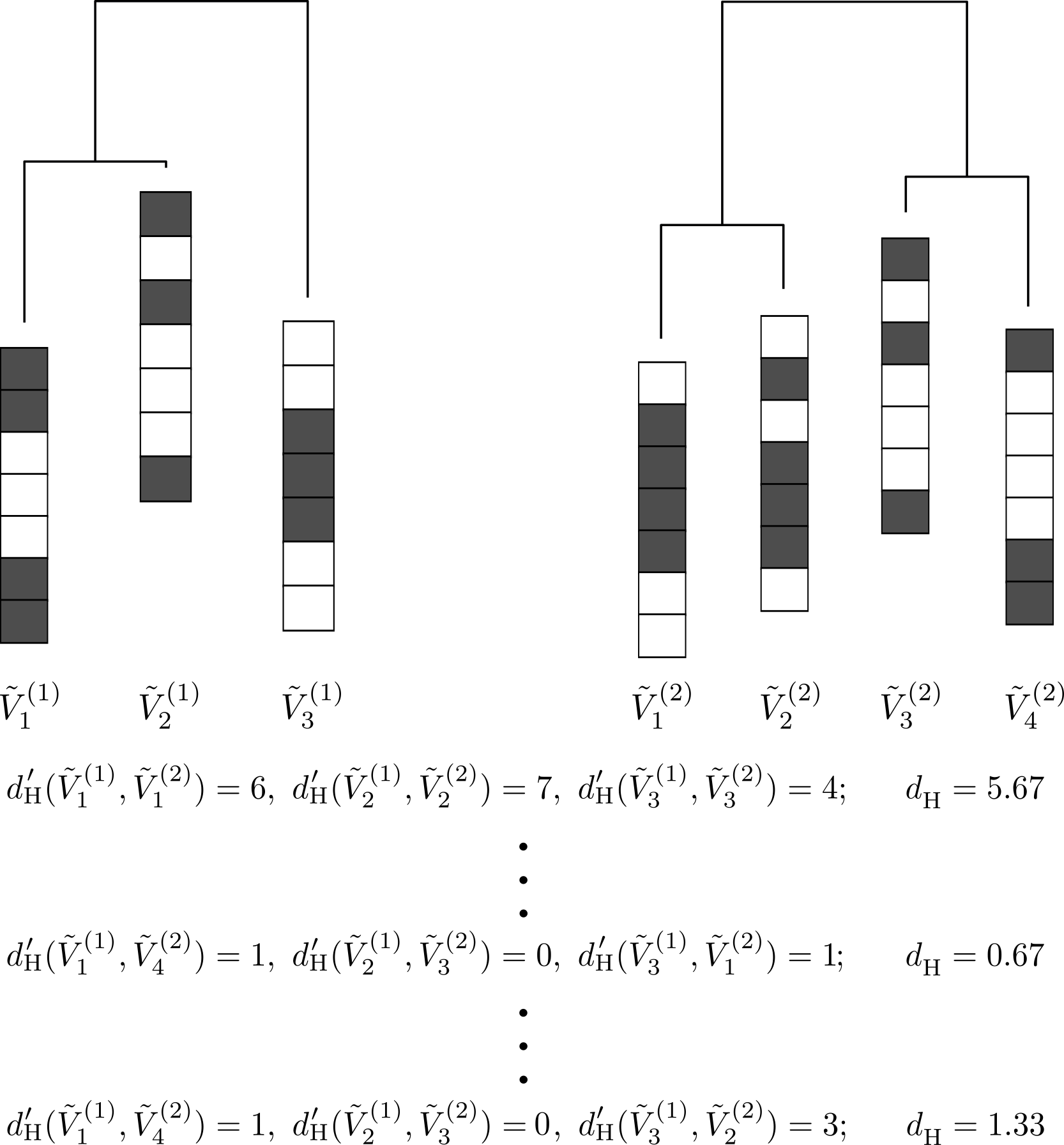}
\caption{Schematic diagram describing how to match the local minima between two energy landscapes. In this example, there are three and four local minima in the first and second energy landscapes, respectively. With the first matching given by $\{(\tilde V_1^{(1)},\tilde V_1^{(2)}),(\tilde V_2^{(1)},\tilde V_2^{(2)}),(\tilde V_3^{(1)},\tilde V_3^{(2)})\}$, we obtain $d_{\text{H}} = 5.67$. We calculate $d_{\text{H}}$ for all the possible $24$ matchings. The smallest $d_{\text{H}}$ value is $0.67$. We adopt the matching that minimizes $d_{\text{H}}$, i.e., $\{(\tilde V_1^{(1)},\tilde V_4^{(2)}), (\tilde V_2^{(1)},\tilde V_3^{(2)}), (\tilde V_3^{(1)},\tilde V_1^{(2)})\}$.
}
\label{Figure_Hamming}
\end{center}
\end{figure}

We denote the activity patterns at the major local minima of the first energy landscape by $\tilde{V}^{(1)}_1$, $\ldots$, $\tilde{V}^{(1)}_{m_1}$, where $m_1$ is the number of the major local minima in the first energy landscape. Similarly, we denote the activity patterns at the major local minima of the second energy landscape by
$\tilde{V}^{(2)}_1$, $\ldots$, $\tilde{V}^{(2)}_{m_2}$. To examine similarity between $\{ \tilde{V}^{(1)}_1, \ldots, \tilde{V}^{(1)}_{m_1} \}$ and $\{ \tilde{V}^{(2)}_1, \ldots, \tilde{V}^{(2)}_{m_2} \}$, we need to match the major local minima between the two energy landscapes. To this end, we assume without loss of generality that $m_1 \le m_2$ and pair each $\tilde{V}^{(1)}_{\ell}$ (with $\ell \in \{1, \ldots, m_1\}$) with a $\tilde{V}^{(2)}_{\ell'}$ (with $\ell' \in \{1, \ldots, m_2\}$) under the condition that different $\tilde{V}^{(1)}_{\ell}$'s are not matched to the same $\tilde{V}^{(2)}_{\ell'}$. We call the obtained correspondence between $\{ \tilde{V}^{(1)}_1, \ldots, \tilde{V}^{(1)}_{m_1} \}$ and $\{ \tilde{V}^{(2)}_1, \ldots, \tilde{V}^{(2)}_{m_2} \}$ a matching. Figure~\ref{Figure_Hamming} describes how to match between the local minima of two energy landscapes.

Note that $m_2 - m_1$ major local minima in the second energy landscape are not matched to any major local minimum in the first energy landscape. We quantify the quality of a matching by
\begin{equation}
d_\text{H} = \frac{1}{m_1}\sum_{\ell=1}^{m_1} d_\text{H}'\left(\tilde{V}^{(1)}_{\ell}, \tilde{V}^{(2)}_{\rho(\ell)}\right),
\end{equation}
where $\tilde{V}^{(2)}_{\rho(\ell)}$ is the activity pattern at the major local minimum paired with $\tilde{V}^{(1)}_{\ell}$ in the considered matching;
$d_\text{H}'$ is the Hamming distance between the $N$-dimensional binary vectors
$\tilde{V}^{(1)}_{\ell}$ and $\tilde{V}^{(2)}_{\rho(\ell)}$, i.e., the number of ROIs whose binary activity (i.e., $\sigma_i = -1$ or $+1$) is opposite between 
$\tilde{V}^{(1)}_{\ell}$ and $\tilde{V}^{(2)}_{\rho(\ell)}$. 
We calculate $d_\text{H}$ for all the possible matchings and select the one that minimizes $d_\text{H}$, which we simply refer to as $d_\text{H}$ hereafter.
A small $d_\text{H}$ value implies that the two energy landscapes are similar in terms of the activity patterns at the local minima of energy.

\subsubsection{Discrepancy in terms of the activity patterns averaged over the attractive basin}
\label{cosine_similarity}

Brain dynamics tend to visit local minima of the energy landscape but also fluctuate around it. Therefore, we additionally measured a distance between the two energy landscapes in terms of the activity patterns averaged over the attractive basin of local minima as follows.

Consider a major local minimum of the first energy landscape, $\tilde{V}^{(1)}_\ell$.
The attractive basin of $\tilde{V}^{(1)}_{\ell}$ is a set of activity patterns. By definition, $V$ is in the attractive basin of 
$\tilde{V}^{(1)}_{\ell}$ if and only if the gradient-descent walk starting from $V$ eventually reaches $\tilde{V}^{(1)}_{\ell}$. The gradient-descent walk on the set of activity patterns is defined by a series of moves from an activity pattern to another such that the move from $V$ is allowed only when the next activity pattern is the one that attains the smallest energy (i.e., largest probability of appearance) among the neighbors of $V$. Intuitively, if we release a ball at $V$, the ball following the gradient moves on the energy landscape until it reaches $\tilde{V}^{(1)}_{\ell}$ and stops there if there is no dynamical noise.

We calculate the average of the activity patterns within the attractive basin of $\tilde{V}^{(1)}_{\ell}$,
which we denote by $\bm{u}^{(1)}_{\ell}$. Note that $\bm{u}^{(1)}_{\ell}$ is an $N$-dimensional vector, which we assume to be a column vector, whose $i$th entry is the average of $\sigma_i \in \{-1, 1\}$ over all the activity patterns in the attractive basin of $\tilde{V}^{(1)}_{\ell}$. 

Similarly, denote by $\bm{u}^{(2)}_{\ell'}$ the average of the activity patterns in the attractive basin of
$\tilde{V}^{(2)}_{\ell'}$ in the second energy landscape.
Then, we calculate the cosine distance between $\bm{u}^{(1)}_{\ell}$ and $\bm{u}^{(2)}_{\ell'}$ given by
\begin{equation}
d_{\rm basin}' \left( \bm{u}^{(1)}_{\ell}, \bm{u}^{(2)}_{\ell'} \right) = 1-\frac{\bm{u}^{(1)\top}_{\ell} \bm{u}^{(2)}_{\ell'}}
{\left\| \bm{u}^{(1)}_{\ell} \right\| \cdot \left\| \bm{u}^{(2)}_{\ell'} \right\|},
\end{equation}
where $\lVert~\rVert$ denotes the Euclidean norm of the vector. The $d_{\rm basin}'$ value ranges between $0$ and $2$. A small value of $d_{\rm basin}'$ indicates a stronger alignment between $\bm{u}^{(1)}_{\ell}$ and $\bm{u}^{(2)}_{\ell'}$. 
For a given matching $\rho$, we then define
\begin{equation}
d_{\rm basin}=\frac{1}{m_1}\sum_{\ell=1}^{m_1}d_{\rm basin}' \left( \bm{u}^{(1)}_{\ell}, \bm{u}^{(2)}_{\rho(\ell)} \right),
\end{equation}
which quantifies overall discrepancy between the two energy landscapes in terms of the average activity pattern in the attractive basin of the local minimum.
We calculate $d_{\rm basin}$ for all the possible matchings between $\{ \tilde{V}^{(1)}_1, \ldots, \tilde{V}^{(1)}_{m_1} \}$ and $\{ \tilde{V}^{(2)}_1, \ldots, \tilde{V}^{(2)}_{m_2} \}$ and adopt the smallest value, which we also refer to as $d_{\rm basin}$ for simplicity. In a majority of cases, the best matching determined by the minimization of $d_\text{H}$ and that determined by the minimization of $d_{\rm basin}$ are the same. However, they are sometimes different from each other.

\subsubsection{Discrepancy in terms of the branch length}
\label{normalized_branch}

As a fourth measure to characterize energy landscapes, we quantified the ease with which the activity pattern switches from one major local minimum to another. We call it the normalized branch length. Then, we compared the normalized branch length between two energy landscapes.

We compute the normalized branch length as follows. We first calculate the length of the branch corresponding to each major local minimum $\gamma$ as the difference between the energy value of $\gamma$ and the smallest energy value at which $\gamma$ joins the branch of another major local minimum on the disconnectivity graph. The calculated branch length quantifies the difficulty of transitioning from $\gamma$ to another local minimum.

We assume that there are $m_1$ and $m_2$ major local minima from the first and second energy landscapes, respectively. We denote by $L^{(1)}$ and $L^{(2)}$ the average of the branch length over the $m_1$ corresponding branches in the first energy landscape and over the $m_2$ branches in the second energy landscape, respectively. Then, we define the normalized branch length difference between the two energy landscapes by
\begin{equation}
d_L=\frac{\left| L^{(1)}-L^{(2)} \right|}{\max (L^{(1)},L^{(2)})}.
\label{eq:def-d_L}
\end{equation}

\subsection{Nonparametric statistical analysis\label{sub:ND}}

We examine whether the energy landscapes estimated from different fMRI data from the same participants are more similar to each other than the energy landscapes estimated for two different groups of participants. We argue that, if the energy landscape analysis is useful, the energy landscapes estimated from the same participants should be closer to each other than the energy landscapes estimated from different participants.

\begin{figure*}[h]
\begin{center}
\includegraphics[width=17cm]{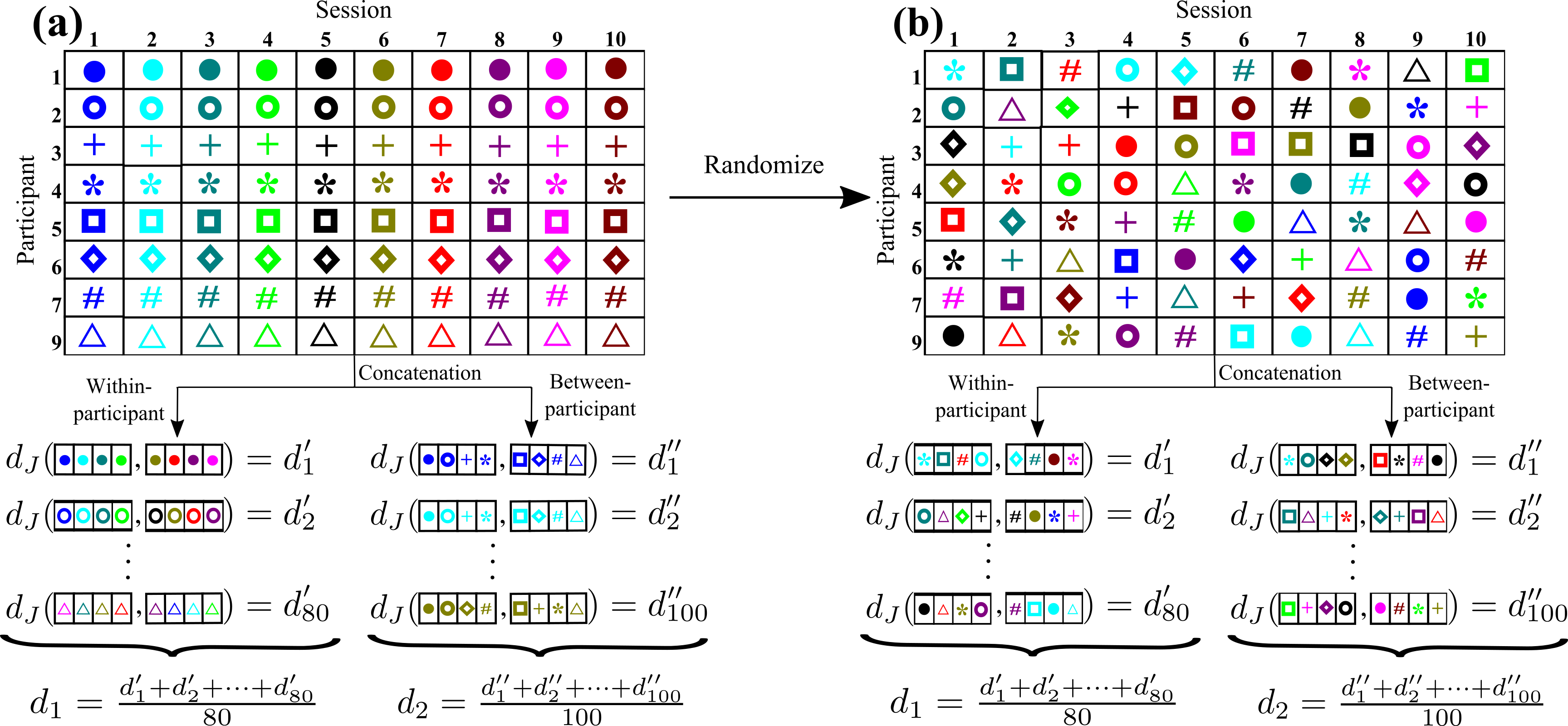}
\caption{Schematic diagram describing concatenation of the fMRI data across different sessions and calculation of $d_1$ and $d_2$. (a) For the original data. (b) For the randomized data. The inference of the energy landscape is based on the data concatenated across four sessions. The same four cells in the table are used for the concatenation in (a) and (b). However, because of the random permutation, any concatenation in (b) is over four sessions that are selected uniformly at random from the original data. Therefore, in (b), $d'_{1}$ and $d''_{1}$, for example, are statistically the same, and the expectation of $d_1$ and that of $d_2$ are the same.
}
\label{ND_schematic}
\end{center}
\end{figure*}

First, we consider the MSC data with the conventional likelihood maximization method, for which we need to concatenate fMRI data over sessions to estimate one energy landscape with a reasonably high accuracy.
For expository purposes, we consider one of the four discrepancy measures, say, $d_J$. We also focus on the $p$th participant. We first calculate $d_J$ between $J^{(1)}$ and $J^{(2)}$, where we estimate $J^{(1)}$ for the fMRI data concatenated over four sessions that are uniformly randomly selected out of the ten sessions, $s \in \{1, \ldots, 10 \}$, and $J^{(2)}$ from the fMRI data concatenated over another uniformly randomly selected four sessions. We impose that the second set of four sessions does not overlap with the first set. Note that we use eight out of ten randomly selected sessions to calculate one $d_J$ value. We repeat this procedure ten times to obtain ten values of $d_J$ for the $p$th participant. By calculating ten values of $d_J$ for each of the eight participants, i.e., $p\in \{1, \ldots, 7, 9\}$, we obtain $8 \times 10=80$ values of $d_J$. We denote the average of the 80 values of $d_J$ by $d_1$ (see Fig.~\ref{ND_schematic}(a)).

Next, we calculate $d_J$, with $J^{(1)}$ being estimated for the fMRI data concatenated over the $s$th sessions of four participants that are uniformly randomly selected out of the eight participants, and $J^{(2)}$ from the fMRI data concatenated over the $s$th sessions of the other four participants. We repeat this procedure ten times to obtain ten values of $d_J$ for the $s$th session. By calculating the ten values of $d_J$ for each of the ten sessions, i.e., $s\in \{ 1, \ldots, 10 \}$, we obtain $10 \times 10=100$ values of $d_J$. We denote the average of the 100 values of $d_J$ by $d_2$ (see Fig.~\ref{ND_schematic}(a)).

Second, we consider the case in which we do not need to concatenate fMRI data before estimating an energy landscape. We again consider $d_J$ as an example. 
We first calculate $d_J$ between $J^{(1)}$ and $J^{(2)}$, where $J^{(i)}, i=\{1,2\}$, is estimated for two sessions $s$ and $s'$ $(\neq s)$ for a participant $p$. It should be noted that $s, s' \in \{1,\ldots,10\}$ and $p \in \{1,\ldots,7,9\}$ for the MSC data, and $s, s' \in \{1,\ldots,4\}$ and $p \in \{1,\ldots, 87\}$ for the HCP data.
By exhausting all pairs $(s, s')$, we compute $n_1$ values of $d_J$, where $n_1= 10 \times 9 / 2 = 45$ and $n_1 = 4 \times 3/2 = 6$ for the MSC and HCP data, respectively, for each participant. We denote by $d_1$ the average of the $8n_1$ or $87n_1$ values of $d_J$ for the MSC or HCP data, respectively, which originate from all pairs $(s, s')$ and all participants.

Next, we calculate $d_J$, with $J^{(1)}$ and $J^{(2)}$ being estimated for the $s$th session for two different participants $p$ and $p'$ $(\neq p)$, where $p, p' \in \{1,\dots,7,9\}$ for the MSC data and $p, p' \in \{1,\dots,87\}$ for the HCP data. By exhausting all pairs of participants $(p, p')$, we compute $n_2$ values of $d_J$, where $n_2=8 \times 7/2=28$ and $n_2 = 87 \times 86/2=3741$ for the MSC and HCP data, respectively, for each session. We denote by $d_2$ the average of the $10n_2$ or $4n_2$ values of $d_J$ for the MSC or HCP data, respectively.

We define ${\rm ND} \equiv d_2/d_1$; ND is named after normalized distance~\cite{Maris_Neuroscience2007nonparametric, Liu_Neuro2020geometric}. If energy landscapes are more similar between different sets of sessions from the same participant (i.e., within-participant comparison) than between those from different participants (i.e., between-participant comparison), the ND value will be larger than $1$. In this case, we regard that the energy landscape analysis bears high within-participant test-retest reliability. In contrast, if the energy landscape from the same participant is not particularly reliable across sessions,
the ND will be close to $1$. 

To statistically examine whether ND is sufficiently larger than $1$, we run a nonparametric permutation test, which is an adaptation of the same test in different studies~\cite{Maris_Neuroscience2007nonparametric, Liu_Neuro2020geometric}. The steps of the permutation test based on the ND are as follows. Here we use $d_J$ to explain the steps. See Fig.~\ref{ND_schematic}(b) for a schematic of randomized MSC data.

\begin{enumerate}

\item Consider the binarized $N$-dimensional fMRI time series data for each of the eight participants and each of the ten sessions.

\item \label{randomize} Uniformly randomly permute the 80 participant-session pairs in the case of the MSC data or 348 ($=87 \times 4$) participant-session pairs in the case of the HCP data.  After the randomization, the fMRI data for the $s$th session from the $p$th participant is the fMRI data for a uniformly randomly selected session from a uniformly randomly selected participant without replacement.

\item \label{randomized_ND} We calculate ND for the randomized data. For the combination of the MSC data and the conventional likelihood maximization method, this step entails concatenating the fMRI data over four random sessions from the same participant $p$ or over the $s$th sessions from four random participants, estimate the energy landscapes for the concatenated data, comparing two energy landscapes to calculate $d_J$, repeat this 80 times to obtain $d_1$ and 100 times to obtain $d_2$, and compute ${\rm ND} = d_2/d_1$. When the data are not concatenated (i.e., the MSC data with the variational Bayes method or the HCP data with the conventional method), we calculate $d_J$ for each pair of sessions from the same participant $p$ and take the average to obtain $d_1$. We also calculate $d_J$ for each pair of participants from each session $s$ and take the average to obtain $d_2$. Then, we set ND$=d_2/d_1$.

\item Repeat steps (\ref{randomize}) and (\ref{randomized_ND}) over $c$ random permutations, where $c$ is a large number. We set $c=10^3$.

\item Calculate the $p$ value, which is the fraction of the random permutations that yield an ND value larger than that for the original data.

\item If the $p$ value is significantly small, then we reject the null hypothesis that $d_1= d_2$ in the original data. In this case, we conclude the significant presence of within-participant test-retest reliability in the energy landscape analysis.
\end{enumerate}

In step~\ref{randomized_ND}, we calculate $d_1$ and $d_2$ as the within-participant and between-participant averages, respectively. However, for the randomized data, they are statistically the same except that $d_1$ and $d_2$ are averages of $80$ and $100$ values of $d_J$, respectively (for the case of the MSC data combined with the conventional likelihood maximization method). This is because $d_J$ calculated for both $d_1$ and $d_2$ originates from the comparison of the energy landscape estimated from uniformly randomly selected four out of the 80 sessions and another uniformly randomly selected four sessions without overlapping. Therefore, $d_1$ and $d_2$ have the same mean, and ND is expected to be peaked approximately at 1. The present permutation test thus evaluates whether the reliability of the energy landscape analysis across sessions for the same participant is higher than that across sessions for different participants.

\section{Results}

\subsection{Accuracy of fit of the pairwise MEM}

We extracted $N$ ROIs for three brain networks, i.e., the whole-brain network ($N=7$), DMN ($N=8$), and CON ($N=7$).
For each of them, we estimated the pairwise MEM for the resting-state fMRI signals obtained from healthy adults in the MSC data set. 

We calculated $r_D$, the accuracy of fit of the pairwise MEM, for each pair of participant and session. We obtained $r_D = 69.12 \pm 6.41\%$ (average $\pm$ standard deviation) for the whole brain network, $r_D = 57.97 \pm 8.94\%$ for the DMN, and $r_D = 77.65 \pm 5.41\%$ for the CON (also see Table~\ref{table:within-participants}). 

Because the accuracy of fit is not high enough, as is customarily done, we concatenated the data across participants or across sessions, estimated the pairwise MEM, and calculated $r_D$~\cite{Watanabe_NatCom2014energy, Ezaki_PTRS2017energy}. Specifically, we concatenated the fMRI data across $m$ sessions, where $m \in \{ 2, 3, 4, 5\}$. The $m$ sessions are from the same participant but from $m$ different sessions, or have the same session ID (i.e., $s$) but from $m$ different participants. We show in Table~\ref{table:within-participants} the average and standard deviation of $r_D$ for the three networks when we concatenated $m \in \{2, 3, 4, 5 \}$ sessions from the same participant. 
Table~\ref{table:between-participants} shows the $r_D$ values when we concatenated $m$ sessions from different participants.
In both Tables~\ref{table:within-participants} and \ref{table:between-participants}, as expected, $r_D$ increases as $m$ increases ($\beta_1 = 3.09$ in Eq.~\eqref{eq:r_D-regression}; $p=4.60 \times 10^{-9}$). 
Furthermore, $r_D$ is larger with the within-participant than across-participant concatenation ($\beta_4 = 3.51$ in Eq.~\eqref{eq:r_D-regression}; $p= 6.04 \times 10^{-5}$). 
The latter result indicates that the energy landscape estimated through the within-participant concatenation of the fMRI data is more accurate than that estimated through the between-participant concatenation in terms of the accuracy of fit of the pairwise MEM.

\begin{table}[t]
\centering
\begin{tabular}{|wc{0.1cm}|wc{1.76cm}|wc{1.76cm}|wc{1.76cm}|}
%\begin{tabular}{|c|c|c|c|}
     \hline
      $m$ & \begin{tabular}{c} Whole-brain \\ network \end{tabular}& DMN & CON\\
      \hline
      1 &  $69.12 \pm 6.41$ & $57.97 \pm 8.94$ & $77.65 \pm 5.41$\\
       \hline
      2 &  $81.26 \pm 4.36$ & $71.49 \pm 5.77$ & $86.26 \pm 3.51$\\
        \hline
      3 & $85.97 \pm 3.34$ & $76.78 \pm 5.30$ & $90.07 \pm 2.32$\\
        \hline
      4 &  $88.62 \pm 3.49$ & $80.80 \pm 4.44$ & $92.30 \pm 1.70$\\
        \hline
      5 &  $90.59 \pm 2.45$ & $83.73 \pm 4.44$ & $93.50 \pm 1.54$\\
        \hline
    \end{tabular}
    \caption{\textbf{Accuracy of fit of the pairwise MEM when we concatenate fMRI data within the same participant.} Each cell represents the accuracy of fit in percent when we concatenate the fMRI data across sessions from the same participant. We concatenated data from a given participant over $m$ sessions and then fitted the pairwise MEM to the concatenated data. With $m=2$, we partitioned the 10 sessions into 5 groups as (1,2), (3,4), (5,6), (7,8), and (9,10), concatenated the fMRI data within each group and within each participant, estimated the energy landscape, and computed the accuracy of fit, $r_D$. For example, we concatenated the data from the first two scanning sessions from participant 1, estimated the energy landscape, and computed $r_D$. We did the same for data from the third and fourth sessions from participant 1, the first and second sessions from participant 2, for example. With $m=3$, we concatenated sessions $s= 1,2$, and $3$ from the same participant into one time series, sessions $s=4,5$ and $6$ into one series, and sessions $s=7, 8$, and $9$ to one series. With $m=4$, we concatenated sessions $s= 1, 2, 3$, and $4$ into one series and sessions $s= 5, 6, 7$, and $8$ into another series. With $m=5$, we concatenated sessions $s=1, 2, 3, 4$, and $5$ into one series and sessions $s=6, 7, 8, 9$, and $10$ into another series. The average and standard deviation were computed across the participants and the different manners to concatenate $m$ sessions per participant.}
\label{table:within-participants}
\end{table}

In both Tables~\ref{table:within-participants} and \ref{table:between-participants}, the accuracy for the DMN is substantially lower than that for the whole-brain network ($\beta_2 = 10.34$ in Eq.~\eqref{eq:r_D-regression}; $p=1.63 \times 10^{-10}$) and the CON ($\beta_3 = 14.31$ in Eq.~\eqref{eq:r_D-regression}; $p=5.54 \times 10^{-13}$). This is presumably because the DMN has one more ROI than the whole-brain network and the CON. The accuracy decreases as the number of ROIs increases in general~\cite{Ezaki_PTRS2017energy}.

\begin{table}[h]
\begin{tabular}{|wc{0.1cm}|wc{1.76cm}|wc{1.76cm}|wc{1.76cm}|}
      \hline
      $m$ & \begin{tabular}{c} Whole-brain \\ network \end{tabular} & DMN & CON\\
      \hline
      1 &  $69.12 \pm 6.41$ & $57.97 \pm 8.94$ & $77.65 \pm 5.41$\\
       \hline
      2 &  $79.64 \pm 3.77$ & $64.71 \pm 7.45$ & $84.81 \pm 3.68$\\
        \hline
      3 & $83.92 \pm 1.96$ & $72.41 \pm 4.51$ & $86.66 \pm 3.13$\\
        \hline
      4 &  $86.36 \pm 1.85$ & $73.46 \pm 3.59$ & $90.76 \pm 2.07$\\
        \hline
      5 &  $87.51 \pm 1.72$ & $77.79 \pm 3.23$ & $91.27 \pm 1.06$\\
        \hline
    \end{tabular}
    \caption{\textbf{Accuracy of fit of the pairwise MEM when we concatenate fMRI data across different participants.} Each cell represents the accuracy of fit in percent when we concatenate the fMRI data across sessions from different participants. We concatenated data from a given session over $m$ participants and then fitted the pairwise MEM to the concatenated data. With $m=2$, we concatenated the data for the $s$th session from participants $p=1$ and $2$ into one time series, those from participants $p=3$ and $4$ into another series, those from participants $p= 5$ and $6$ into another series, and those from participants $p=7$ and $9$ into another series. We did this for each $s$. With $m=3$, we concatenated the data from participants $p= 1,2$, and $3$ into one series and those from participants $p=4, 5$, and $6$ into another series. With $m=4$, we concatenated the data from participants $p= 1, 2, 3$, and $4$ into one series and those from participants $p=5,6,7$, and $9$ into another series. With $m=5$, we concatenated the data from participants $p= 1, 2, 3, 4$, and $5$ into one series. Note that the results with $m=1$ shown in this table are identical with those in Table~\ref{table:within-participants}. The average and standard deviation were computed across the sessions and the different manners to concatenate $m$ participants given the session.}
\label{table:between-participants}
\end{table} 

In the following analyses, we use concatenation over $m=4$ sessions and examine test-retest reliability of the energy landscape analysis. Figure~\ref{ND_schematic}(a) schematically explains the concatenation within each participant and that across participants. With $m=4$, the accuracy of fit is more than 85\% except for the DMN. In general, we are also interested in the test-retest reliability of fMRI data in the case of a relatively low accuracy of fit, which we test with the DMN. A concatenation over more sessions, such as with $m=5$, would further increase the accuracy of fit (see Tables~\ref{table:within-participants} and \ref{table:between-participants}). Then, however, examining test-retest reliability may be more difficult because one needs to create two energy landscapes, preferably from non-overlapping data, and systematically compare them. In the present study, we use data obtained from eight participants. Therefore, if $m=5$, one cannot avoid overlapping of the participants if we create two groups of participants for concatenating the fMRI data. Our choice of $m=4$ balances the accuracy of fit and the tractability of the test-retest reliability analysis.

\begin{figure}[H]
\begin{center}
\includegraphics[width=8.8cm]{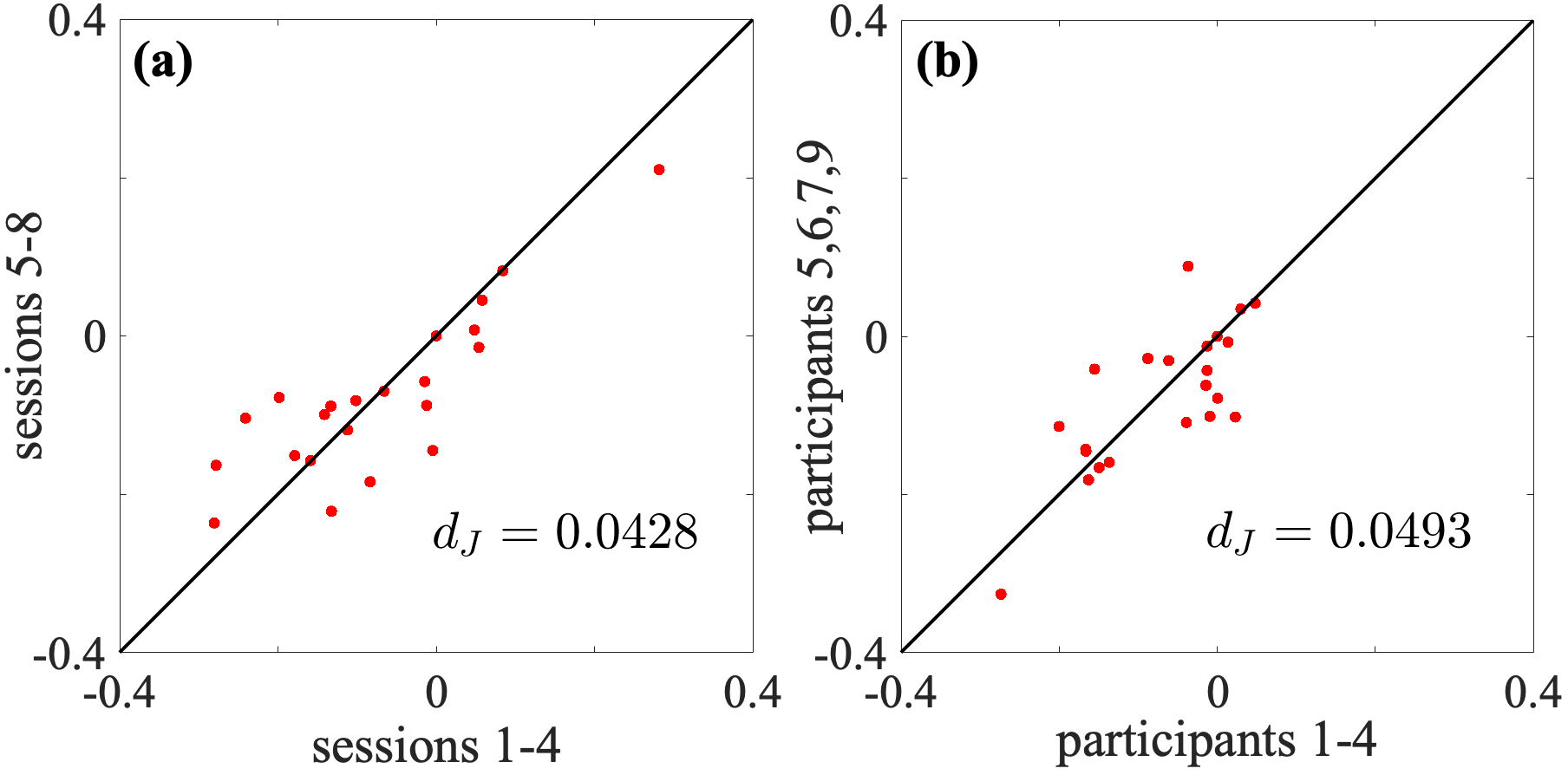}
\caption{\textbf{Reliability of the interaction strength between two ROIs, $J_{ij}$, for the whole-brain network.} (a) Within-participant comparison. We concatenated the data for the first participant over four sessions. The horizontal and vertical axes correspond to the concatenation of sessions 1 to 4 and sessions 5 to 8, respectively. Each circle represents a pair of $i$ and $j$. (b) Between-participant comparison. We concatenated the data from the first session over four participants. The horizontal and vertical axes correspond to the concatenation of the first and last four participants, respectively. In both (a) and (b), if all the circles lay on the diagonal, which we show by the solid lines, then the discrepancy, $d_J$, would be equal to 0. The $d_J$ value is large if the circles tend to be far from the diagonal.
}
\label{fig:Discrepancy}
\end{center}
\end{figure}

\begin{table*}[t]
\centering
\setlength{\tabcolsep}{7pt}
\begin{tabular}{|m{0.9cm}|m{2.23cm}|m{2.23cm}|m{2.23cm}|m{2.23cm}|m{2.23cm}|m{2.23cm}|}
     \hline
      & \multicolumn{2}{c|}{Whole-brain network} & \multicolumn{2}{c|}{DMN} & \multicolumn{2}{c|}{CON}\\
      \hline
      &  Within ($d_1$) & Between ($d_2$) & Within ($d_1$) & Between ($d_2$) & Within ($d_1$) & Between ($d_2$)\\
      \hline
      $d_J$ &  $0.0464 \pm 0.0082$ & $0.0527 \pm 0.0098$ & $0.0448 \pm 0.0091$ & $0.0634 \pm 0.0060$ & $0.0432 \pm 0.0130$ & $0.0684 \pm 0.0110$\\
       \hline
      $d_\text{H}$ & $0.5800 \pm 1.0956$ & $0.9125 \pm 0.8149$ & $0.5417 \pm 0.4777$ & $1.7333 \pm 1.0998$ & $0.3854 \pm 0.4606$ & $0.6667 \pm 0.3909$\\
        \hline
      $d_{\rm basin}$ & $0.0232 \pm 0.0239$ & $0.0346 \pm 0.0329$ & $0.0245 \pm 0.0220$ & $0.0578 \pm 0.0245$ & $0.0157 \pm 0.0098$ & $0.0236 \pm 0.0121$\\
        \hline
      $d_L$ & $0.2852 \pm 0.2100$ & $0.3529 \pm 0.1625$ & $0.2022 \pm 0.2366$ & $0.3526 \pm 0.3341$ & $0.2281 \pm 0.1679$ & $0.3671 \pm 0.1683$\\
        \hline
\end{tabular}
\caption{Discrepancy between two energy landscapes estimated by the conventional likelihood maximization applied to the MSC data. ``Within'' and ``Between'' in the table stand for within-participant and between-participant comparison, respectively. We computed the average and standard deviation of $d_1$ and $d_2$ across the participants and across the sessions, respectively.}
\label{table:Conventional_results}
\end{table*}

\begin{table}[ht]
\centering
\setlength{\tabcolsep}{9.5pt}
\begin{tabular}{|wl{0.7cm}|wc{1.78cm}|wc{1.78cm}|wc{1.78cm}|}
    \hline
    & \begin{tabular}{c} Whole-brain \\ network \end{tabular} & DMN & CON\\
    \hline
    $d_J$ & \begin{tabular}{c} $\text{ND} = 1.315$ \\  $p<10^{-3}$ \end{tabular} & \begin{tabular}{c} $\text{ND} = 1.415$ \\ $p<10^{-3}$ \end{tabular} & \begin{tabular}{c} $\text{ND} = 1.580$ \\ $p<10^{-3}$ \end{tabular} \\
    \hline
    $d_\text{H}$ & \begin{tabular}{c} $\text{ND} = 1.934$ \\ $p<10^{-3}$ \end{tabular} & \begin{tabular}{c} $\text{ND} = 3.200$ \\ $p<10^{-3}$ \end{tabular} & \begin{tabular}{c} $\text{ND} = 1.730$ \\ $p<10^{-3}$ \end{tabular} \\
    \hline
    $d_{\rm basin}$ & \begin{tabular}{c} $\text{ND} = 1.491$ \\ $p<10^{-3}$ \end{tabular} & \begin{tabular}{c} $\text{ND} = 2.359$ \\ $p<10^{-3}$ \end{tabular} & \begin{tabular}{c} $\text{ND} = 1.503$ \\ $p = 0.003$ \end{tabular} \\
    \hline
    $d_L$ & \begin{tabular}{c} $\text{ND} = 1.237$ \\ $p = 0.014$ \end{tabular} & \begin{tabular}{c} $\text{ND} = 1.744$ \\ $p<10^{-3}$ \end{tabular} & \begin{tabular}{c} $\text{ND} = 1.609$ \\ $p<10^{-3}$ \end{tabular} \\
    \hline
\end{tabular}
\caption{ND values and the permutation test results for the four discrepancy measures, calculated with the conventional likelihood maximization applied to the MSC data. The $p$ values are the uncorrected values.}
\label{table:Conventional_NDpermutation}
\end{table}

\subsection{Reliability in terms of the interaction strength\label{sub:result-conventional-1}}

We first examined the test-retest reliability of the energy landscape analysis in terms of the interaction strength parameters $\{J_{ij}\}$. We concatenated the fMRI data over the first four sessions from the $p$th participant and estimated $\{J_{ij}\}$ for each $p \in \{ 1, 2, 3, 4, 5, 6, 7, 9\}$.
Similarly, for each participant $p$, we concatenated the data over the next four sessions (i.e., sessions 5 to 8) and estimated $\{J_{ij}\}$.
For the whole-brain network, we show the relationships between $J_{ij}$ estimated for the first four sessions against that estimated for the next four sessions for the first participant in Fig.~\ref{fig:Discrepancy}(a). Each circle represents $J_{ij}$ for a pair of $i$ and $j$. The values of $\{J_{ij}\}$ are reasonably consistent between the first four sessions and the next four sessions (Pearson correlation coefficient $= 0.850$; discrepancy $d_J = 0.0428$).

We instead concatenated the data for a single session over the first four participants (i.e., $p=1$, $2$, $3$, and $4$) and estimate $\{J_{ij}\}$, did the same for the last four participants (i.e., $p=5$, $6$, $7$, and $9$), and compared the two obtained sets of $\{J_{ij}\}$. In this manner, we investigated the consistency of the energy landscape between participants. For the whole-brain network, we show relationships between $\{J_{ij}\}$ for the two sets of participants in the first session in Fig.~\ref{fig:Discrepancy}(b). Similar to the case of Fig.~\ref{fig:Discrepancy}(a), the estimated $\{J_{ij}\}$ was reasonably consistent between the two concatenations, consistent with previous results with other data~\cite{Watanabe_NatCom2013pairwise, Ezaki_HBM2018, Kang_Neuroimage2017energy}. However, the degree of consistency was smaller for the present between-participant comparison (Pearson correlation coefficient $= 0.773$; discrepancy $d_J = 0.0493$) than the within-comparison comparison. In this particular example, the estimation of $\{J_{ij}\}$ was more consistent between pairs of sessions from the same participant than those from different participants.
 
To examine the generality of this result, we then calculated $d_J$ between the concatenation across sessions 1 to 4 and that across sessions 5 to 8 from the same participant (i.e., within-participant comparison). The mean and standard deviation of $d_J$ over the eight participants were equal to $d_J = 0.0464 \pm 0.0082$ (mean $\pm$ std) for the whole-brain network. We also calculated $d_J$ between the concatenation of the $s$th section over the first four participants and that over the last four participants (i.e., between-participant comparison). The mean and standard deviation of $d_J$ for the between-participant comparison over the ten sessions were equal to  $d_J = 0.0527  \pm 0.0098$. We show these $d_J$ values and those for the DMN and CON in Table~\ref{table:Conventional_results}. The table suggests that the energy landscape is apparently somewhat more similar between different fMRI sessions obtained from the same participant than between different participants.

To statistically investigate potential differences between the within-participant and between-participant comparisons, we carried out the permutation test on $d_J$. The $\text{ND}$ for the whole-brain network, DMN, and CON were at least $1.3$ (see Table~\ref{table:Conventional_NDpermutation}). After a random permutation of the participants and sessions, the ND value was centered around 1 by definition. We show the distribution of the ND value obtained from $c=10^3$ random permutations in Fig.~\ref{fig:ND_MSC}(a), (b), and (c) for the whole-brain network, DMN, and CON, respectively. We calculated the $p$ value for the empirical data by contrasting it to the distribution of ND for the randomized data. We obtained $p<10^{-3}$ for all the three networks, implying that no random permutation yielded an ND value larger than that for the empirical data before the random permutation. These results remained significant after correction for the multiple comparison present in Table~\ref{table:Conventional_NDpermutation} ($p<1.2 \times 10^{-2}$, Bonferroni corrected). Therefore, we conclude that the estimated parameter values, $\{J_{ij}\}$, are significantly more reliable in the within-participant than between-participant comparison for the three networks.

\subsection{Reliability in terms of the activity patterns at the local minima}

As a second index of the consistency between different energy landscapes, we compared the activity patterns at the local minima of the energy landscape between energy landscape pairs in terms of the Hamming distance, $d_\text{H}$. Table~\ref{table:Conventional_results} indicates that the average $d_\text{H}$
is  at least 1.6 times larger for the between-participant than within-participant comparison for the whole-brain network, DMN, and CON.

\begin{figure}[t]
\begin{center}
\includegraphics[width=8.8cm]{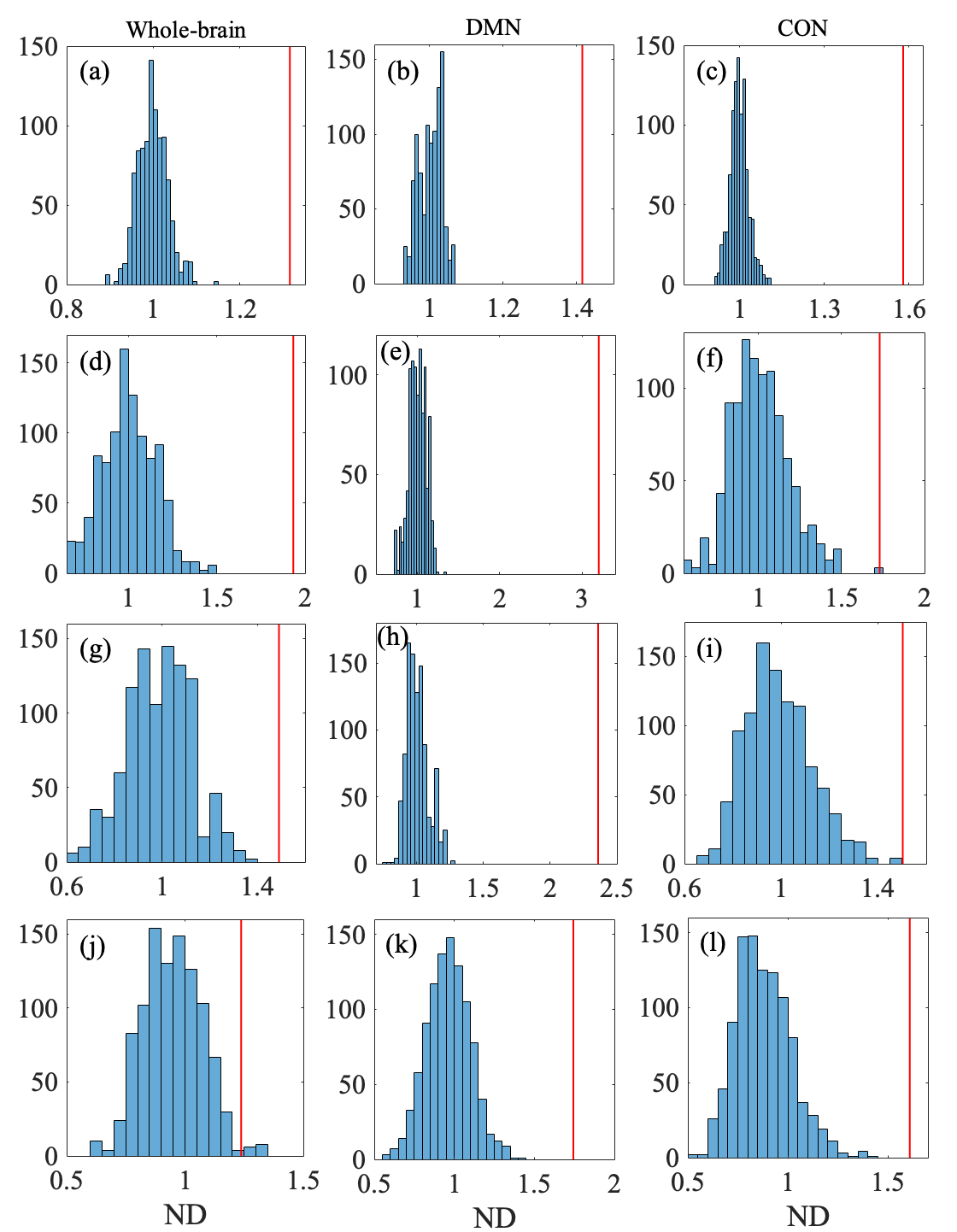}
\caption{\textbf{Histogram of ND for the randomized data and the empirical ND value.} The first, second, and third columns of the figure show the distributions for the whole-brain network, DMN, and CON, respectively. The four rows of the figure show the distributions for $d_J$ (in (a), (b), and (c)), $d_\text{H}$ (in (d), (e), and (f)), $d_{\rm basin}$ (in (g), (h), and (i)), and $d_L$ (in (j), (k), and (l)), from the top to the bottom. In each panel, the vertical line indicates the empirical ND value.}
\label{fig:ND_MSC}
\end{center}
\end{figure}

The $\text{ND}$ value was at least $1.73$ for the three networks (see Table~\ref{table:Conventional_NDpermutation}).
The permutation test yielded $p < 10^{-3}$ for all the three networks; see Fig. \ref{fig:ND_MSC}(d)--(f) for the distribution of the ND values for the random permutations.
These results altogether support that the reliability of the energy landscape analysis in terms of $d_\text{H}$ is higher within the same participant than between different participants.

\subsection{Reliability in terms of the activity patterns averaged over the attractive basin}

As a third index to characterize the consistency between energy landscapes, we measured the distance between the average activity patterns belonging to the attractive basin of a local minimum in one energy landscape and that in another energy landscape, i.e., $d_{\rm basin}$.
Similarly to the case of $d_J$ and $d_\text{H}$, we found that $d_{\rm basin}$ is substantially smaller for the within-participant than between-participant comparison for the three networks although the standard deviation is not small (see Table~\ref{table:Conventional_results}). It should be noted that the observed $d_{\rm basin}$ values are close to $0$ for both the within-participant and between-participant comparisons. This result implies the almost full agreement between a pair of energy landscapes in terms of the averaged activity pattern in the attractive basin, even for the between-participant comparison.

We show in Fig.~\ref{fig:ND_MSC}(g)--(i) for the distribution of the ND values for the random permutations as well as the ND values for the original energy landscapes. 
The permutation test yielded $p < 10^{-3}$ for the whole-brain network and the DMN and $p=0.003$ for the CON (Table~\ref{table:Conventional_NDpermutation}).
These results support a significantly high test-retest reliability of the energy landscape analysis in terms of $d_{\rm basin}$ including the case of the CON after correction for multiple comparisons across the networks and indices ($p = 0.036$, Bonferroni corrected).

\subsection{Reliability in terms of the branch length\label{sub:result-conventional-4}}

As a last index of consistency of the energy landscape, we measure the normalized difference in the average branch length in the disconnectivity graph, $d_L$, between two energy landscapes. We found that the average of $d_L$ was smaller for the within-participant than between-participant comparison for the three networks (see Table~\ref{table:Conventional_results}). The permutation test yielded $p=0.014$ for the whole-brain network, and $p < 10^{-3}$ for the DMN and CON; see Table~\ref{table:Conventional_NDpermutation} and Fig.~\ref{fig:ND_MSC}(j)--(l). These results support a significantly high test-retest reliability of the energy landscape analysis in terms of $d_L$
for the DMN and CON although the result for the whole-brain network did not survive correction for multiple comparison ($p=0.17$, Bonferroni corrected).

\subsection{Accuracy and reliability of the variational Bayes approximation method}

The Bayesian estimation potentially allows us to reliably estimate an energy landscape without concatenating fMRI data across sessions or participants even if a single session is not long. Therefore, we repeated the same test-retest reliability analysis on the MSC data with the Bayesian estimation and without any concatenation.

After running the variational Bayes approximation method to compute the hyperparameters, we calculated the accuracy of fit, $r_D$, of the pairwise MEM. We obtained $r_D = 86.02 \pm 2.79\%$, $91.50 \pm 3.21\%$, and $93.51 \pm 1.48\%$ for the whole-brain network, DMN, and CON, respectively. These high accuracy values support the effectiveness of the method.
 
We show the mean and standard deviation of the four discrepancy indices for the within-participant and between-participant comparison in Table~\ref{table:Bayesian_results}. For some combinations of the session, participant, and network, the Bayesian method yielded an energy landscape with just one local (and hence global) minimum of the energy. In this case, we set the branch length to be $0$. Table~\ref{table:Bayesian_results} suggests that the within-participant consistency of energy landscape analysis is notably higher than the between-participant consistency in terms of the four discrepancy measures although the standard deviation is large. These results are qualitatively the same as those obtained with the conventional likelihood maximization method described in sections~\ref{sub:result-conventional-1}--\ref{sub:result-conventional-4}. However, the discrepancy values are substantially larger with the Bayesian method (see Table~\ref{table:Bayesian_results}) than the likelihood maximization method (see Table~\ref{table:Conventional_results}) for both within-participant and between-participant comparisons with few exceptions. 

\begin{table*}[t]
\centering
\setlength{\tabcolsep}{7pt}
\begin{tabular}{|m{0.9cm}|c|c|c|c|c|c|}
     \hline
      & \multicolumn{2}{c|}{Whole-brain network} & \multicolumn{2}{c|}{DMN} & \multicolumn{2}{c|}{CON}\\
      \hline
      &  Within ($d_1$) & Between ($d_2$) & Within ($d_1$) & Between ($d_2$) & Within ($d_1$) & Between ($d_2$)\\
      \hline
      $d_J$ &  $0.2748 \pm 0.0693$ & $0.3500 \pm 0.0741$ & $0.3291 \pm 0.1501$ & $0.4455 \pm 0.1352$ & $0.2482 \pm 0.0743$ & $0.3428 \pm 0.0766$\\
       \hline
      $d_\text{H}$ & $1.1038 \pm 0.5951$ & $1.4910 \pm 0.5846$ & $1.5439 \pm 1.0098$ & $2.3617 \pm 0.8919$ & $0.7881 \pm 0.6890$ & $1.2365 \pm 0.7095$\\
        \hline
      $d_{\rm basin}$ & $0.0582 \pm 0.0305$ & $0.0696 \pm 0.0272$ & $0.0342 \pm 0.0262$ & $0.0436 \pm 0.0257$ & $0.0211 \pm 0.0160$ & $0.0285 \pm 0.0172$\\
        \hline
      $d_L$ & $0.3537 \pm 0.2076$ & $0.3765 \pm 0.2066$ & $0.5056 \pm 0.3198$ & $0.5342 \pm 0.3120$ & $0.2610 \pm 0.2051$ & $0.3036 \pm 0.2057$\\
        \hline
\end{tabular}
\caption{Discrepancy between two energy landscapes estimated by the variational Bayes method applied to the MSC data.
``Within'' and ``Between'' in the table stand for within-participant and between-participant, respectively. We computed the average and standard deviation of $d_1$ and $d_2$ across the participants and across the sessions, respectively.}
\label{table:Bayesian_results}
\end{table*}

Table~\ref{table:Bayesian_NDpermutation} shows the results of the permutation test for the three networks and four discrepancy measures. We find significantly higher reliability within the same participant than between different participants in terms of $d_J$, $d_H$ and $d_{\rm basin}$. In terms of $d_L$, the uncorrected $p$ values were smaller than $0.05$ but did not survive the Bonferroni correction for the whole-brain network and DMN. These results were similar to those for the likelihood maximization method. However, comparison of Tables~\ref{table:Conventional_NDpermutation} and \ref{table:Bayesian_NDpermutation} reveals that the ND value with the Bayesian method was smaller than that with the likelihood maximization method for all the four discrepancy measures and all the three networks.
Therefore, we conclude that the Bayesian method yields significantly higher reliability within the same participant than between different participants in most cases, whereas the reliability is somewhat weaker than in the case of the conventional likelihood maximization method.

\begin{table}[t]
\centering
\setlength{\tabcolsep}{9.5pt}
\begin{tabular}{|wl{0.7cm}|wc{1.78cm}|wc{1.78cm}|wc{1.78cm}|}
% \begin{tabular}{|m{0.7cm}|c|c|c|}
     \hline
       & \begin{tabular}{c} Whole-brain \\ network \end{tabular} & DMN & CON\\
      \hline
      $d_J$ &  \begin{tabular}{c} $\text{ND} = 1.274$ \\ $p<10^{-3}$ \end{tabular} & \begin{tabular}{c} $\text{ND} = 1.354$ \\ $p<10^{-3}$ \end{tabular} & \begin{tabular}{c} $\text{ND} = 1.381$ \\ $p<10^{-3}$ \end{tabular} \\
       \hline
      $d_\text{H}$ & \begin{tabular}{c} $\text{ND} = 1.351$ \\ $p<10^{-3}$ \end{tabular} & \begin{tabular}{c} $\text{ND} = 1.530$ \\ $p<10^{-3}$ \end{tabular} & \begin{tabular}{c} $\text{ND} = 1.569$ \\ $p<10^{-3}$ \end{tabular} \\
        \hline
      $d_{\rm basin}$ & \begin{tabular}{c} $\text{ND} = 1.196$ \\ $p<10^{-3}$ \end{tabular} & \begin{tabular}{c} $\text{ND} = 1.275$ \\ $p<10^{-3}$ \end{tabular} & \begin{tabular}{c} $\text{ND} = 1.351$ \\ $p<10^{-3}$ \end{tabular} \\
        \hline
      $d_L$ & \begin{tabular}{c} $\text{ND} = 1.065$ \\ $p = 0.0320$ \end{tabular} & \begin{tabular}{c} $\text{ND} = 1.057$ \\ $p = 0.0230$ \end{tabular} & \begin{tabular}{c} $\text{ND} = 1.163$ \\ $p<10^{-3}$ \end{tabular} \\
        \hline
    \end{tabular}
    \caption{ND values and the permutation test results for the four discrepancy measures, calculated with the variational Bayes method applied to the MSC data. The $p$ values are the uncorrected values.
    }
\label{table:Bayesian_NDpermutation}
\end{table}

\subsection{Validation with the Human Connectome Project data}

\begin{table}[t]
\centering
\setlength{\tabcolsep}{11pt}
\begin{tabular}{|m{0.7cm}|c|c|c|}
     \hline
      &  Within ($d_1$) & Between ($d_2$)\\
      \hline
      $d_J$ &  $0.0784 \pm 0.0194$ & $0.1027 \pm 0.0232$\\
       \hline
      $d_\text{H}$ &  $0.3145 \pm 0.4864$ & $0.3623 \pm 0.4967$\\
        \hline
      $d_{\rm basin}$ & $0.0309 \pm 0.0314$ & $0.0386 \pm 0.0325$\\
        \hline
      $d_L$ &  $0.2535 \pm 0.1619$ & $0.2921 \pm 0.1783$\\
        \hline
\end{tabular}
\caption{Discrepancy between two energy landscapes estimated by the conventional likelihood maximization method applied to the HCP data. ``Within'' and ``Between'' in the table stand for within-participant and between-participant, respectively. We computed the average and standard deviation of $d_1$ and $d_2$ across the participants and across the sessions, respectively.}
\label{table:HCP_results}
\end{table}

As a different type of validation, we ran the test-retest reliability analysis for another fMRI data set, HCP data. 
We used a whole-brain network with $N=7$ ROIs.
We calculated the accuracy of fit, $r_D$, of the pairwise MEM estimated with the likelihood maximization method to single-session data. We obtained $r_D = 92.49 \pm 1.99\%$, where we calculated the average and standard deviation on the basis of the four sessions per participant and all the participants.
Table~\ref{table:HCP_results} shows the mean and standard deviation of the four discrepancy indices, compared between the within-participant and between-participant comparison. The results are similar to those for the MSC data. The ND values for $d_J$, $d_\text{H}$, $d_{\rm basin}$, and $d_L$ are $1.310$, $1.152$, $1.249$, and $1.152$, respectively. The permutation test yielded $p < 10^{-3}$ for all the four discrepancy indices. These results confirm significantly higher within-participant than between-participant test-retest reliability of the energy landscape analysis with a different data set.

\subsection{Permutation test by shuffling the participants within each session}

As another validation, we carried out a different variant of the permutation test in which we shuffled the participants within each session; the same shuffling was employed in previous studies~\cite{Termenon_Neuroimage2016reliability, Wang_HBM2017test}. Most combinations of the discrepancy measure and the network showed significantly small $p$ values after the Bonferroni correction for the different data sets and energy landscape inference methods. (See Appendix B~\ref{Appendix_B} for the detailed methods and results.) This result is similar to the case of those with our original shuffling method schematically shown in Fig.~\ref{ND_schematic}.

\section{Discussion}

We examined test-retest reliability of the energy landscape analysis in terms of four indices. For each index, we calculated a discrepancy in the index value between two estimated energy landscapes. We then constructed and ran a permutation test on the calculated discrepancy value to statistically assess whether within-participant comparison of two energy landscapes yielded a smaller discrepancy value than between-participant comparison of two energy landscapes. For the two data sets, we found significant within-participant test-retest reliability (i.e., within-participant discrepancy being significantly smaller than between-participant discrepancy) in most cases. Furthermore, we found qualitatively the same results for a Bayesian variant of the energy landscape estimation method that enables us to estimate an energy landscape for each scanning session, mitigating the data-hungry nature of the original estimation method.

The accuracy of fit measured by $r_D$ was large for the variational Bayes approximation method (i.e., $86.02$, $91.50$, and $93.51$\% on average for the whole-brain network, DMN, and CON, respectively) although we did not concatenate the fMRI data across different sessions. These $r_D$ values are close to that with the conventional likelihood maximization method with concatenation of four or five sessions (see Tables~\ref{table:within-participants} and \ref{table:between-participants}). The high accuracy of the variational Bayes method is presumably due to the fact that the target empirical distribution of activity patterns, i.e., $P_N(V_i)$ in Eq.~\eqref{eq:D_ell-def}, is necessarily different between the two estimation methods. Specifically, $P_N(V_i)$ is the empirical distribution over all sessions for non-Bayesian estimation methods, whereas it is the empirical distribution for one session for Bayesian methods. We do not ascribe the higher accuracy of fit of the variational Bayes method to overfitting. The variational Bayes method yields a Boltzmann distribution for each session. Therefore, it uses $M\times D$ parameters, where we remind that $M=N(N+1)/2$ is the number of parameters of the Boltzmann distribution and $D=80$ is the number of sessions. Therefore, it uses $D$ times more parameters than the conventional likelihood maximization method, which uses $M$ parameters to estimate one Boltzmann distribution. However, the variational Bayes method needs to produce an accurate Boltzmann distribution tailored to a single session to attain a high $r_D$ value, which is not the case for the conventional likelihood maximization method. In general, the accuracy of the pairwise MEM simply degrades if the data are shorter (see Tables~\ref{table:within-participants} and \ref{table:between-participants}; also see \cite{Ezaki_PTRS2017energy} for a systematic analysis on the effect of the data length on the accuracy). Our results that the variational Bayes method yields a higher accuracy of fit and higher consistency in the within-participant than between-participant comparison both support that individual-to-individual differences are not negligible when carrying out energy landscape analysis. While such individual differences were a motivation behind the original proposals of the Bayesian methods~\cite{Kang_HBM2021bayesian,Jeong_Neuroimage2021empirical}, further comparisons of Bayesian and non-Bayesian estimation methods as well as pursuit of biological and medical relevances of energy landscapes estimated with the Bayesian methods remain future work. 

The significance of the test-retest reliability results obtained with the permutation test was similar between the likelihood maximization and variational Bayes methods. However, the ND values were larger for the likelihood maximization than the variational Bayes method. As a separate result, the discrepancy indices were overall smaller for the likelihood maximization than Bayesian method. The latter two results are in favor of the likelihood maximization over Bayesian method for realizing high test-retest reliability. However, we point out that the estimation of an energy landscape for the likelihood maximization requires concatenation of four sessions, whereas the Bayesian method avoids concatenation. Assessment of test-retest reliability for different Bayesian approximation methods~\cite{Jeong_Neuroimage2021empirical} and other approximate methods such as the pseudo-likelihood maximization~\cite{Ezaki_PTRS2017energy, Ezaki_CommBio2020closer}, including systematic analysis on the dependence of the results on the data length, is left as future work.

The intraclass correlation coefficient (ICC) has been widely used for investigating test-retest reliability in functional connectivity data~\cite{Noble_Neuroimage2019decade}. We did not use the ICC because our quantification of the estimated energy landscape was mostly multidimensional and difficult to fit to an ANOVA or similar framework based on which most ICC measures are calculated. Specifically, $\{J_{ij}\}$, based on which we calculated $d_J$, is a $N(N-1)/2$-dimensional vector. In addition, we calculated $d_{\text{H}}$ and $d_{\text{basin}}$ by examining the activity patterns at local minima and their average over the attractive basin, respectively, in the situation where the number of the local minima varies in one energy landscape from another. Therefore, we decided to calculate a discrepancy measure for each of the four indices between two energy landscapes and constructed a permutation test to examine test-retest reliability. We point out that the average branch length is a scalar characterization of an energy landscape, and therefore it is straightforward to use conventional ICC measures if we discard the normalization factor in Eq.~\eqref{eq:def-d_L}. See below for a preliminary analysis of ICCs.

Quantities $d_1$ and $d_2$ used for defining ND ($= d_2/d_1$) are averages over 80 and 100 samples, respectively, of a discrepancy measure, such as $d_J$. For the randomized data produced in the permutation test, averaging over 80 or 100 samples kills fluctuations in individual samples. Therefore, the standard deviations of $d_1$ and $d_2$ are small compared to if they were calculated as averages over fewer samples of randomized data. Then, the statistical fluctuation of ND is proportionately small such that ND for the randomized data is centered around $1$ with a small standard deviation, which tends to make the ND for the original data significantly different from $1$. The number of samples for calculating $d_1$ or $d_2$, such as 80 or 100, is our arbitrary choice, and the statistical significance of the permutation test depends on the choice of these numbers. This is an important limitation of our permutation test. However, for the present fMRI data, we still obtain qualitatively similar, albeit statistically weaker, results even without carrying out any averaging. We show in Appendix C~\ref{Appendix_C} the ND values for the original data and the $p$ values from the permutation test when $d_1$ and $d_2$ are a single sample of a discrepancy measure (e.g., $d_J$). Tables~\ref{table:New_Conventional_NDpermutation} and \ref{table:New_Bayesian_WholeBrain} indicate that three out of the twelve combinations of the network and the discrepancy measure yield significant $p$ values (i.e., less than $0.05/12 = 0.00417$ uncorrected, considering the Bonferroni correction) for the MSC data when we estimate the energy landscape by the conventional method and the variational Bayes method, respectively. Table~\ref{table:New_Conventional_HCP} also suggests that, out of the four discrepancy measures calculated for the HCP data with the conventional method, one measure yields a significant $p$ value (i.e., less than $0.05/4=0.0125$ uncorrected).

As another stress test, we briefly analyzed three measures of ICCs although we have already stated why we did not use them in our main analysis. As mentioned earlier, the average branch length, used for defining $d_L$, is the only scalar characterization of an energy landscape employed in the present study. Therefore, we computed the average branch length for the whole-brain network obtained from each session of the HCP data and then the first two ICC measures, which are conventional and take a scalar value for each session as input. The third ICC measure assumes vector input for each session, so we use the vectorization of matrix $J=(J_{ij})$. We show the definition of the three ICC measures in 
Appendix D. The ICC value calculated by a standard method \cite{Koo_ChiMed2016guideline} is 
$-0.0115$. In general, negative values of the ICC are interpreted as zero reliability~\cite{Bartko_PsyReport1966intraclass, Yokum_ACN2021test}.
The ICC value calculated for the average branch length by a second method \cite{Noble_BehavSci2021guide} is $0.4542$. This value is reasonably large \cite{Noble_BehavSci2021guide}. We also calculated $\rm{I_{diff}}$ \cite{Amico_SciReport2018quest, Chiem_Brain2022improving} as an ICC measure. We obtained $\rm{I_{diff}} = 10.13\%$ when the data for each participant-session pair is $\{J_{ij} ; 1 \le i < j \le N\}$. This value is roughly similar to the $\rm{I_{diff}}$ values for functional networks obtained from the HCP data in a previous study \cite{Amico_SciReport2018quest}. The results for the last two ICC measures indicate a moderate reliability within a single participant relative to across different participants. In contrast, the first definition of the ICC does not support this result. Although reasons for the discrepancy are unclear, we believe that these preliminary results are a step to in-depth individual-level fingerprinting analyses in the future.

The estimated $J$ matrix can be regarded as a functional connectivity matrix and can be better at estimating the structural connectivity than other conventional definitions of functional connectivity \cite{Watanabe_NatCom2013pairwise}. The estimated $h_1$, $\ldots$, $h_N$ are close to zero because our standard choice of the threshold (i.e., time average of the signal for each ROI $i$) makes each $\sigma_i$ to take $-1$ and $+1$ approximately with probability $1/2$ each. Then, an energy landscape is almost completely determined by $J$. In this sense, the reliability of the energy landscapes in terms of the indices we have investigated is caused by the reliability of functional connectivity. However, our results are not direct consequences of known results of high reliability of functional connectivity in fMRI data~\cite{Finn_NatNeuro2015functional, Miranda_PlosOne2014connectotyping, Chiem_Brain2022improving} because they estimated functional connectivity using other conventional methods such as the Pearson correlation coefficient and their variants. Furthermore, which properties of energy landscapes, including $J$, bear higher reliability is not a trivial question. For example, Table~\ref{table:New_Conventional_NDpermutation} shows that the activity patterns at the local minimum of the energy, measured by $d_{\text{H}}$, and the activity patterns averaged over the attractive basin, measured by $d_{\rm basin}$, are more reliable than $J$ for the DMN.

Related to this issue, although we proposed four discrepancy indices for pairs of energy landscapes, they are our arbitrary choices. One can apply the analysis pipeline proposed in the present study to assess test-retest reliability for other discrepancy indices. Other potential discrepancy indices are the frequency of transiting from one particular local minimum to another and features of the transition probability matrix among the activity patterns or among the local minima. Furthermore, our framework of the permutation test on the ND value is not limited to energy landscape analysis (e.g., application to ``microstate dynamics'' for fMRI data~\cite{Islam_BMC2024state}).

Individual variability of fMRI data has most frequently been investigated in terms of functional connectivity~\cite{Noble_Neuroimage2019decade}. In contrast, we have shown evidence that energy landscape analysis of fMRI data bears session-to-session reproducibility within a participant relative to between different participants. The present results encourage further work toward application of energy landscape analysis to identification of individuals in different cognitive, behavioral, and clinical conditions.

\section{Acknowledgements}

T.W. acknowledges support from the Japan Society for Promotion of Sciences (TW, 19H03535, 21H05679, 23H04217) 
N.M. acknowledges support from the Japan Science and Technology Agency (JST) Moonshot R\&D (under grant no.\,JPMJMS2021), the National Science Foundation (under grant no.\,2204936), and JSPS KAKENHI (under grant nos.\,JP 21H04595 and 23H03414).

\section{Data availability statement}

The two data sets used in this work are publically available. The first data set was provided by the Midnight Scan Club (MSC) project, funded by NIH Grants NS088590, TR000448 (NUFD), MH104592 (DJG), and HD087011 (to the Intellectual and Developmental Disabilities Research Center at Washington University); the Jacobs Foundation (NUFD); the Child Neurology Foundation (NUFD); the McDonnell Center for Systems Neuroscience (NUFD, BLS); the Mallinckrodt Institute of Radiology (NUFD); the Hope Center for Neurological Disorders (NUFD, BLS, SEP); and Dart Neuroscience LLC. This data was obtained from the OpenfMRI database. Its accession number is ds000224. The second data set was provided by the Human Connectome Project, WU-Minn Consortium (Principal Investigators: David Van Essen and Kamil Ugurbil; 1U54MH091657) funded by the 16 NIH Institutes and Centers that support the NIH Blueprint for Neuroscience Research; and by the McDonnell Center for Systems Neuroscience at Washington University.

\section{Ethical approval}

All the data were open to the public and collected with the approval of the ethics committees or institutional review board of the recording institutes in the reference~\cite{Van_Neuroimage2013wu, Gordon_Neuron2017precision}. The corresponding author's institutional review board generally assigns a determination of Not Human Research to such a case.

\section{Author's contribution}

TW and NM conceptualized the project. JN, SM, and TW collected the data. PK, JN, SM and TW curated the data. PK, TW, and NM developed the methodology. PK performed formal analysis and investigation. PK and NM validated the results. PK, and NM visualized the results. TW and NM collected the funding. NM administered and supervised the project. PK and NM prepared the first draft of the manuscript and other authors contributed to the manuscript revision. All authors read and approved the final version of the manuscript.

\section{Appendix A: Sample disconnectivity graphs \label{Sample_disconnectivity}}

\begin{figure}[H]
\begin{center}
\includegraphics[width=8.8cm]{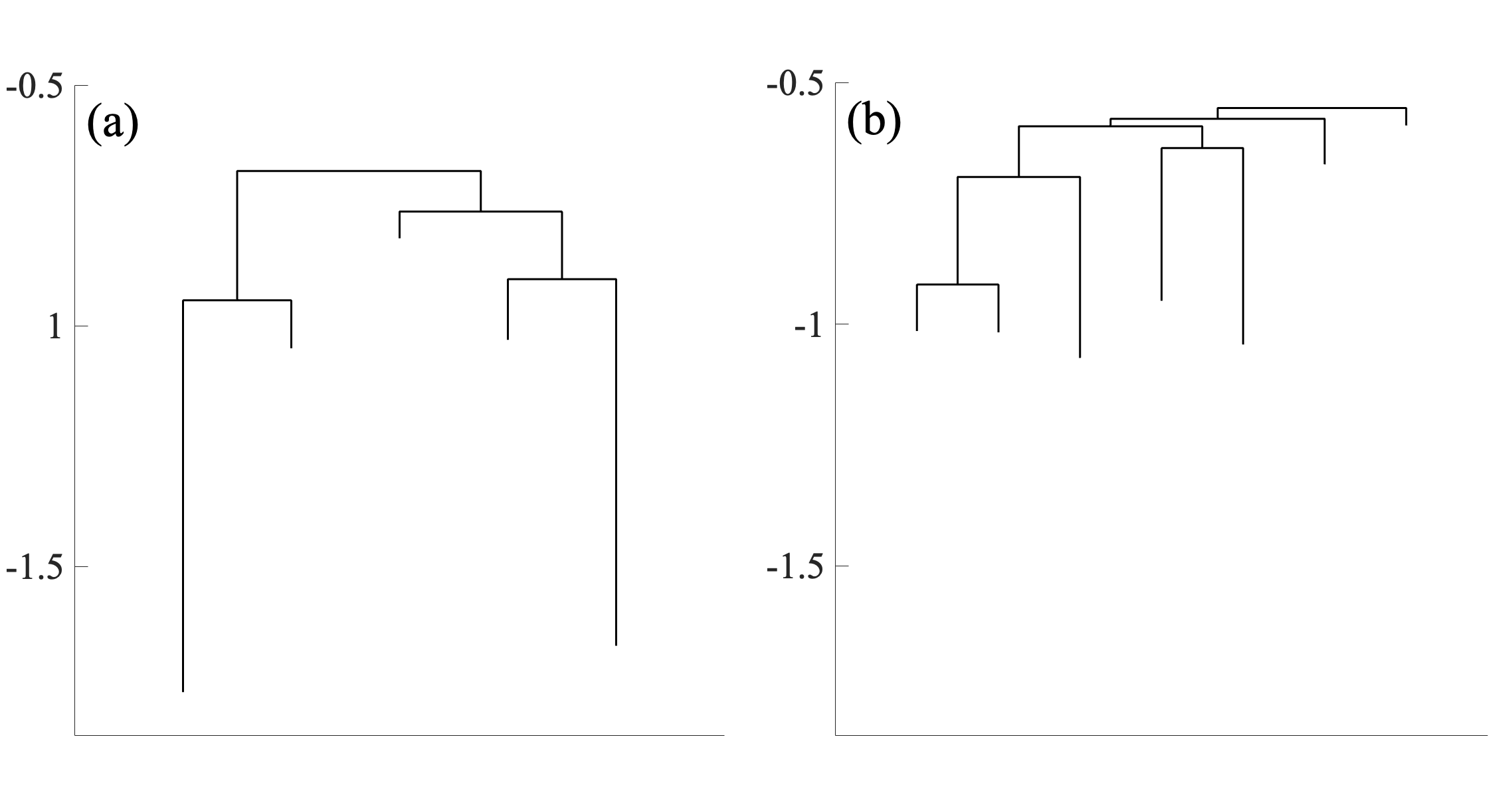}
\caption{Examples of disconnectivity graphs. (a) Concatenation of the first to the fourth sessions from participant MSC01. (b) Concatenation of the four sessions from the same participant after we randomly shuffle the sessions of the MSC data using the method depicted in Fig.~\ref{ND_schematic}. We used the conventional likelihood maximization method.}
\label{Disconnectivity_graph_Review}
\end{center}
\end{figure}

We show a sample disconnectivity graph for the MSC data based on the concatenation of four sessions of a participant the concatenation of four randomized sessions in Fig.~\ref{Disconnectivity_graph_Review}(a) and (b), respectively.

\section{Appendix B: Permutation test by shuffling the participants within each session \label{Appendix_B}}

In this section, we run the permutation test using a different randomization procedure~\cite{Termenon_Neuroimage2016reliability, Wang_HBM2017test}. We first calculate $d_1$ and $d_2$ for the original data in the same way as described in Fig.~\ref{ND_schematic}(a) and then the ND value. However, when randomizing the data for the permutation test, we uniformly randomly permute the fMRI data across participants but within each session. For this randomized data, we calculate $d_1$ and $d_2$, and thus ND. We repeat the randomization followed by the calculation of the ND value $c=10^3$ times to obtain the $p$ value.

We show the ND values for the original data and the $p$ values from the permutation test
for the MSC data combined with the conventional likelihood maximization method, MSC data combined with the variational Bayes method, and HCP data combined with the conventional method in Tables~\ref{table:Conventional_NDpermutation_AcrossParticipants}, \ref{table:Bayesian_NDpermutation_AcrossParticipants}, and \ref{table:Conventional_NDpermutation_HCP_AcrossParticipants}, respectively. We find that the ND values for all the four discrepancy measures and all the three networks are significantly larger than $1$ after the Bonferroni correction for the MSC data when combined with the conventional method (i.e., uncorrected $p<0.05/12 = 0.00417$; see Table~\ref{table:Conventional_NDpermutation_AcrossParticipants}). When combined with the variational Bayes method, 9 out of the 12 combinations of the discrepancy measure and network yield ND values significantly larger than $1$ (see Table~\ref{table:Bayesian_NDpermutation_AcrossParticipants}). The other three  combinations still yield $p<0.05$ before the Bonferroni correction. For the HCP data combined with the conventional method, all the four discrepancy measures are significantly larger than $1$ after the Bonferroni correction (i.e., uncorrected $p<0.05/4 = 0.0125$; Table~\ref{table:Conventional_NDpermutation_HCP_AcrossParticipants}).

\begin{table}[H]
\centering
\setlength{\tabcolsep}{9.5pt}
\begin{tabular}{|wl{0.7cm}|wc{1.78cm}|wc{1.78cm}|wc{1.78cm}|}
%\begin{tabular}{|m{0.7cm}|c|c|c|}
     \hline
       & \begin{tabular}{c} Whole-brain \\ network \end{tabular} & DMN & CON\\
      \hline
      $d_J$ & \begin{tabular}{c} $\text{ND} = 1.315$ \\ $p<10^{-3}$ \end{tabular} & \begin{tabular}{c} $\text{ND} = 1.415$ \\ $p<10^{-3}$ \end{tabular} & \begin{tabular}{c} $\text{ND} = 1.580$ \\ $p<10^{-3}$ \end{tabular} \\
       \hline
      $d_\text{H}$ & \begin{tabular}{c} $\text{ND} = 1.934$ \\ $p<10^{-3}$ \end{tabular} & \begin{tabular}{c} $\text{ND} = 3.200$ \\ $p<10^{-3}$ \end{tabular} & \begin{tabular}{c} $\text{ND} = 1.730$ \\ $p<10^{-3}$ \end{tabular} \\
        \hline
      $d_{\rm basin}$ & \begin{tabular}{c} $\text{ND} = 1.491$ \\ $p<10^{-3}$ \end{tabular} & \begin{tabular}{c} $\text{ND} = 2.359$ \\ $p<10^{-3}$ \end{tabular} & \begin{tabular}{c} $\text{ND} = 1.503$ \\ $p<10^{-3}$ \end{tabular} \\
        \hline
      $d_L$ & \begin{tabular}{c} $\text{ND} = 1.237$ \\ $p = 0.003$ \end{tabular} & \begin{tabular}{c} $\text{ND} = 1.744$ \\ $p = 0.001$ \end{tabular} & \begin{tabular}{c} $\text{ND} = 1.609$ \\ $p<10^{-3}$ \end{tabular} \\
        \hline
    \end{tabular}
    \caption{ND values and the permutation test results, calculated with the conventional method applied to the MSC data and the data randomly shuffled across participants but within each session. The $p$ values are the uncorrected values.}
\label{table:Conventional_NDpermutation_AcrossParticipants}
\end{table}

\begin{table}[H]
\centering
\setlength{\tabcolsep}{9.5pt}
% \begin{tabular}{|m{0.7cm}|c|c|c|}
\begin{tabular}{|wl{0.7cm}|wc{1.78cm}|wc{1.78cm}|wc{1.78cm}|}
     \hline
       & \begin{tabular}{c} Whole-brain \\ network \end{tabular} & DMN & CON\\
      \hline
      $d_J$ & \begin{tabular}{c} $\text{ND} = 1.274$ \\ $p<10^{-3}$ \end{tabular} & \begin{tabular}{c} $\text{ND} = 1.354$ \\ $p<10^{-3}$ \end{tabular} & \begin{tabular}{c} $\text{ND} = 1.381$ \\ $p<10^{-3}$ \end{tabular} \\
       \hline
      $d_\text{H}$ & \begin{tabular}{c} $\text{ND} = 1.351$ \\ $p<10^{-3}$ \end{tabular} & \begin{tabular}{c} $\text{ND} = 1.530$ \\ $p<10^{-3}$ \end{tabular} & \begin{tabular}{c} $\text{ND} = 1.569$ \\ $p<10^{-3}$ \end{tabular} \\
        \hline
      $d_{\rm basin}$ & \begin{tabular}{c} $\text{ND} = 1.196$ \\ $p<10^{-3}$ \end{tabular} & \begin{tabular}{c} $\text{ND} = 1.275$ \\ $p<10^{-3}$ \end{tabular} & \begin{tabular}{c} $\text{ND} = 1.351$ \\ $p=0.012$ \end{tabular} \\
        \hline
      $d_L$ & \begin{tabular}{c} $\text{ND} = 1.065$ \\ $p=0.014$ \end{tabular} & \begin{tabular}{c} $\text{ND} = 1.057$ \\ $p=0.034$ \end{tabular} & \begin{tabular}{c} $\text{ND} = 1.163$ \\ $p<10^{-3}$ \end{tabular} \\
        \hline
    \end{tabular}
    \caption{ND values and the permutation test results, calculated with the variational Bayes method applied to the MSC data and the data randomly shuffled across participants but within each session. The $p$ values are the uncorrected values.}
\label{table:Bayesian_NDpermutation_AcrossParticipants}
\end{table}

\begin{table}[H]
\centering
\setlength{\tabcolsep}{9.5pt}
\begin{tabular}{|wc{1.5cm}|wc{1.5cm}|wc{1.5cm}|wc{1.5cm}|}
     \hline
       $d_J$ & $d_\text{H}$ & $d_{\rm basin}$ & $d_L$\\
      \hline
      $\text{ND} = 1.310$ & $\text{ND} = 1.152$ & $\text{ND} = 1.249$ & $\text{ND} = 1.152$\\
      $p<10^{-3}$ & $p<10^{-3}$ & $p<10^{-3}$ & $p<10^{-3}$\\
       \hline
    \end{tabular}
    \caption{ND values and the permutation test results, calculated with the conventional method applied to the HCP data and the data randomly shuffled across participants but within each session. The $p$ values are the uncorrected values.}
\label{table:Conventional_NDpermutation_HCP_AcrossParticipants}
\end{table}

\section{Appendix C: Permutation test when $d_1$ and $d_2$ are calculated on the basis of one sample\label{Appendix_C}}

In this section, we run the same permutation test for the MSC and HCP data, but when $d_1$ and $d_2$ are just one sample of the discrepancy measure (e.g., $d_J$). Therefore, for the original data, we only used one participant to calculate $d_1$ and one session (e.g., the first session from each participant) to calculate $d_2$. We show the corresponding $\text{ND}$ values for the original data and the $p$ values obtained from the permutation test in Table~\ref{table:New_Conventional_NDpermutation}. 
We applied the conventional likelihood maximization method to the MSC data.
We find that the $\text{ND}$ value is significantly larger than $1$ after the Bonferroni correction (i.e., uncorrected $p < 0.05/12 = 0.00417$) for $d_\text{H}$ and $d_{\rm basin}$ in the DMN and $d_J$ in the CON.

\begin{table}[H]
\centering
\setlength{\tabcolsep}{9.5pt}
% \begin{tabular}{|m{0.7cm}|c|c|c|}
\begin{tabular}{|wl{0.7cm}|wc{1.78cm}|wc{1.78cm}|wc{1.78cm}|}
     \hline
       & \begin{tabular}{c} Whole-brain \\ network \end{tabular} & DMN & CON\\
      \hline
      $d_J$ & \begin{tabular}{c} $\text{ND} = 1.404$ \\ $p=0.341$ \end{tabular} & \begin{tabular}{c} $\text{ND} = 1.320$ \\ $p=0.386$ \end{tabular} & \begin{tabular}{c} $\text{ND} = 1.288$ \\ $p=0.004$ \end{tabular} \\
       \hline
      $d_\text{H}$ & \begin{tabular}{c} $\text{ND} = 1.518$ \\ $p=0.093$ \end{tabular} & \begin{tabular}{c} $\text{ND} = 2.606$ \\ $p<10^{-3}$ \end{tabular} & \begin{tabular}{c} $\text{ND} = 1.440$ \\ $p=0.167$ \end{tabular} \\
        \hline
      $d_{\rm basin}$ & \begin{tabular}{c} $\text{ND} = 1.332$ \\ $p=0.210$ \end{tabular} & \begin{tabular}{c} $\text{ND} = 2.5257$ \\ $p<10^{-3}$ \end{tabular} & \begin{tabular}{c} $\text{ND} = 1.259$ \\ $p=0.222$ \end{tabular} \\
        \hline
      $d_L$ & \begin{tabular}{c} $\text{ND} = 1.015$ \\ $p=0.292$ \end{tabular} & \begin{tabular}{c} $\text{ND} = 1.4165$ \\ $p=0.020$ \end{tabular} & \begin{tabular}{c} $\text{ND} = 1.3769$ \\ $p=0.061$ \end{tabular} \\
        \hline
    \end{tabular}
    \caption{ND values and the permutation test results when we obtain $d_1$ and $d_2$ from just one sample of the discrepancy measure value, calculated with the conventional method applied to the MSC data. The $p$ values are the uncorrected values.
    }
\label{table:New_Conventional_NDpermutation}
\end{table}

We show in Table~\ref{table:New_Bayesian_WholeBrain} the corresponding results for the variational Bayes method applied to the MSC data. We find that the $\text{ND}$ value is significantly larger than $1$ after the Bonferroni correction for $d_H$ in the DMN and CON, and $d_L$ in the CON.

\begin{table}[H]
\centering
\setlength{\tabcolsep}{9.5pt}
%\begin{tabular}{|m{0.7cm}|c|c|c|}
\begin{tabular}{|wl{0.7cm}|wc{1.78cm}|wc{1.78cm}|wc{1.78cm}|}
     \hline
       & \begin{tabular}{c} Whole-brain \\ network \end{tabular} & DMN & CON\\
      \hline
      $d_J$ & \begin{tabular}{c} $\text{ND} = 1.067$ \\ $p=0.317$ \end{tabular} & \begin{tabular}{c} $\text{ND} = 1.314$ \\ $p=0.052$ \end{tabular} & \begin{tabular}{c} $\text{ND} = 1.439$ \\ $p=0.005$ \end{tabular} \\
       \hline
      $d_\text{H}$ & \begin{tabular}{c} $\text{ND} = 1.022$ \\ $p=0.438$ \end{tabular} & \begin{tabular}{c} $\text{ND} = 1.641$ \\ $p=0.002$ \end{tabular} & \begin{tabular}{c} $\text{ND} = 2.246$ \\ $p=0.002$ \end{tabular} \\
        \hline
      $d_{\rm basin}$ & \begin{tabular}{c} $\text{ND} = 1.002$ \\ $p=0.441$ \end{tabular} & \begin{tabular}{c} $\text{ND} = 1.019$ \\ $p=0.091$ \end{tabular} & \begin{tabular}{c} $\text{ND} = 1.001$ \\ $p=0.448$ \end{tabular} \\
        \hline
      $d_L$ &  \begin{tabular}{c} $\text{ND} = 1.1591$ \\ $p=0.305$ \end{tabular} & \begin{tabular}{c} $\text{ND} = 1.2401$ \\ $p=0.1540$ \end{tabular} & \begin{tabular}{c} $\text{ND} = 2.9438$ \\ $p=0.002$ \end{tabular} \\
        \hline
    \end{tabular}
    \caption{ND values and the permutation test results when we obtain $d_1$ and $d_2$ from just one sample of the discrepancy measure value, calculated with the variational Bayes method applied to the MSC data. The $p$ values are the uncorrected values.
    }
\label{table:New_Bayesian_WholeBrain}
\end{table}

We show the results for the conventional likelihood maximization method applied to the HCP data in Table~\ref{table:New_Conventional_HCP}. We find that ND value is significantly larger than $1$ after the Bonferroni correction only for $d_J$.

\begin{table}[H]
\centering
\setlength{\tabcolsep}{9.5pt}
\begin{tabular}{|wc{1.5cm}|wc{1.5cm}|wc{1.5cm}|wc{1.5cm}|}
     \hline
       $d_J$ & $d_\text{H}$ & $d_{\rm basin}$ & $d_L$\\
      \hline
      $\text{ND} = 1.563$ & $\text{ND} = 3.382$ & $\text{ND} = 1.028$ & $\text{ND} = 1.197$\\
      $p=0.002$ & $p=0.141$ & $p=0.140$ & $p=0.321$\\
       \hline
    \end{tabular}
    \caption{ND values and the permutation test results when we obtain $d_1$ and $d_2$ from just one sample of the discrepancy measure value, calculated with the conventional method applied to the HCP data. The $p$ values are the uncorrected values.
    }
\label{table:New_Conventional_HCP}
\end{table}

\section{Appendix D: ICC measures \label{ICC_measures}}

We computed the following three ICC measures for the whole-brain network obtained from the HCP data.

We denote by $L_{p,s}$ the average branch length for the $s$th session of the $p$th participant, where $p \in \{1, \ldots, N_{\text{p}}\}$, $s \in \{1, \ldots, N_{\text{s}} \}$, $N_{\text{p}} = 87$ is the number of participants, and $N_{\text{s}} = 4$ is the number of sessions per participant. We calculate the first two ICC measures for the average branch length. We interpret $(L_{p,s})$ as an $N_{\text{p}} \times N_{\text{s}}$ matrix, in which each row represents a participant and each column represents a session. As the first ICC measure, we use~\cite{Koo_ChiMed2016guideline}
\begin{equation}
{\rm ICC}=\frac{{\rm MS_R}-{\rm MS_E}}{{\rm MS_R}+\frac{{\rm MS_C}-{\rm MS_E}}{N_{\text{p}}}}.
\label{eq:ICC-def-1}
\end{equation}
In Eq.~\eqref{eq:ICC-def-1}, ${\rm MS_R} = \text{SS}_{\text{B}}/(N_{\text{p}}-1)$ is the mean square for rows,
$\text{SS}_{\text{B}}$ is the variability between participants (also known as the sum of squares between participants),
$\text{MS}_{\text{E}} = \text{SS}_{\text{E}}/(N_{\text{p}}N_{\text{s}} -1)$ is the mean square for error,
$\text{SS}_{\text{E}}$ is the residual variability for error (also known as the sum of squares for error),
$\text{MS}_{\text{C}} = \text{SS}_{\text{C}}/(N_{\text{s}}-1)$ is the mean square for columns, and
$\text{SS}_{\text{C}}$ is the variability between sessions (also known as the sum of squares for columns).
We calculate $\text{SS}_{\text{B}}$, $\text{SS}_{\text{E}}$, and $\text{SS}_{\text{C}}$
using ExpDes, which is an R package for ANOVA and experimental designs~\cite{Ferreira_AM2014expdes}.

To calculate the second ICC, we start by calculating the mean of the average branch length over four sessions for each participant $p$ as follows:
\begin{equation}
L(p) = \frac{1}{N_{\text{s}}}\sum_{s=1}^{N_{\text{s}}} L_{p,s}.
\end{equation}
We also calculate the mean of $L_{p,s}$ over all the $N_{\text{p}} N_{\text{s}}$ participant-session pairs as follows:
\begin{equation}
\langle L_{p, s} \rangle = \frac{1}{N_{\text{p}} N_{\text{s}}} \sum_{p=1}^{N_{\text{p}}} \sum_{s=1}^{N_{\text{s}}} L_{p, s}.
\end{equation}
The between-participant variability, denoted by $\sigma_{\rm{participant}}^2$, is given by
\begin{equation}
\sigma^2_{\text{participant}} = \frac{1}{N_{\text{p}}} \sum_{p=1}^{N_{\text{p}}} \left[ L(p)- \langle L_{p,s} \rangle \right]^2.
\end{equation}
The ICC is given by~\cite{Noble_BehavSci2021guide}
\begin{equation}
\text{ICC} = \frac{\sigma_{\rm{participant}}^2}{\sigma_{\rm{total}}^2},
\end{equation}
where $\sigma_{\rm{total}}^2$ is the total variance over all the $N_{\text{p}} N_{\text{s}}$ participant-session pairs and given by
\begin{equation}
\sigma_{\text{total}}^2 = \frac{1}{N_{\text{p}} N_{\text{s}}}\sum_{p=1}^{N_{\text{p}}} \sum_{s=1}^{N_{\text{s}}} (L_{p,s}- \langle L_{p,s} \rangle)^2.
\label{sigma_total}
\end{equation}
The ICC ranges from $0$ to $1$ and is larger when the different participants are more distinct from each other in terms of $L_{p,s}$.

The third ICC measure that we use is defined in \cite{Amico_SciReport2018quest, Chiem_Brain2022improving} as follows. Let $\overline{A}$ be the so-called identifiability matrix. The $(i, j)$ element of the $N_{\text{p}} \times N_{\text{p}}$ matrix $\overline{A} = (\overline{A}_{ij})$ is the average Pearson correlation coefficient between the $i$th and $j$th participant. Specifically, to calculate $\overline{A}_{ii}$, we first calculate the Pearson correlation coefficient between vector $\{ J_{ij} ; 1 \le i < j \le N\}$ for the $s$th session of the $i$th participant and the same vector for the $s'$th session of the $i$th participant, where $s, s' \in \{ 1, \ldots, 4 \}$ and $s \neq s'$. Then, we average the calculated correlation coefficient over all the $4\times 3/2 = 6$ session pairs $(s, s')$, which gives $\overline{A}_{ii}$. To calculate $\overline{A}_{ij}$ for $i\neq j$, we first calculate the Pearson correlation coefficient between the $s$th session of the $i$th participant and the $s'$th session of the $j$th participant for each $s, s' \in \{ 1, \ldots, 4 \}$. Then, we average the calculated correlation coefficient over all the $4\times 4 = 16$ session pairs $(s, s')$, which gives $\overline{A}_{ij}$. Next, we refer to $I_{\text{self}} \equiv \sum_{i=1}^{N_{\text{p}}} \overline{A}_{ii}/N_{\text{p}}$ as self identifiability. Similarly,  $I_{\text{others}} = \sum_{\substack{i,j=1 \\ i < j}}^{N_{\text{p}}} \overline{A}_{ij}/[N_{\text{p}}(N_{\text{p}}-1)/2]$ represents the average of the off-diagonal elements of the identifiability matrix. We define the differential identifiability by
\begin{equation}
I_{\text{diff}} \equiv(I_{\text{self}} - I_{\text{others}}) \times 100.
\end{equation}

\bibliographystyle{elsarticle-num}
\bibliography{Energy_Landscape_master_bib}
\end{document}